\documentclass[twocolumn,tighten]{aastex63}
\usepackage{CJKutf8}
\usepackage{times}
\usepackage{amsmath}
\usepackage{graphicx}
\usepackage{subfigure}
\usepackage{placeins}
\usepackage{hyperref}
\usepackage{gensymb}
\usepackage{upgreek}
\usepackage{multirow}
\usepackage[normalem]{ulem}
\newcommand{\kms}{km~s$^{\mathrm{-1}}$}
\newcommand{\HI}{H{\sevenrm\,I}}
\newcommand{\MHI}{$M_{\mathrm{H\, \textsc{i}}}$}
\newcommand{\mstar}{$M_{*}$}
\newcommand{\msun}{$M_{\mathrm{\odot}}$}

 \font\sevenrm=cmr7 scaled 1000

\received{\today}
\revised{xxxx , 2022}
\accepted{xxxx , 2022}
\submitjournal{ApJS}

\begin{document}
\begin{CJK*}{UTF8}{gbsn} %%%%%%%% add Chinese

\title{Statistical Analysis of H\ \textsc{i} Profile Asymmetry and Shape for Nearby Galaxies}

\correspondingauthor{Jing Wang}
\email{jwang\_astro@pku.edu.cn}

\shorttitle{\HI\, Shape and Asymmetry}
\shortauthors{YU ET AL.}

\author[0000-0002-9066-370X]{Niankun Yu (余捻坤)}
\affiliation{Kavli Institute for Astronomy and Astrophysics, Peking University, Beijing 100871, China}
\affiliation{Department of Astronomy, School of Physics, Peking University, Beijing 100871, China}

\author[0000-0001-6947-5846]{Luis C. Ho}
\affiliation{Kavli Institute for Astronomy and Astrophysics, Peking University, Beijing 100871, China}
\affiliation{Department of Astronomy, School of Physics, Peking University, Beijing 100871, China}
  
\author[0000-0002-6593-8820]{Jing Wang}
\affiliation{Kavli Institute for Astronomy and Astrophysics, Peking University, Beijing 100871, China}
\affiliation{Department of Astronomy, School of Physics, Peking University, Beijing 100871, China}

\author{Hangyuan Li}
\affiliation{Kavli Institute for Astronomy and Astrophysics, Peking University, Beijing 100871, China}

\begin{abstract}

We present a uniform analysis of the integrated profile of the \HI\ emission line of 29,958 nearby ($z < 0.06$) galaxies extracted from the ALFALFA 21~cm survey.  We apply the curve-of-growth technique to derive a database of spectral parameters and robust estimates of their associated uncertainties. Besides the central velocity and total flux, the main catalog provides new measures of line width, profile asymmetry, and profile shape.  For a subsample of 13,511 galaxies with optical properties available from the Sloan Digital Sky Survey, we compute inclination angle-corrected line widths, rotation velocities empirically calibrated from spatially resolved observations, and dynamical masses based on \HI\ sizes estimated from the \HI\ mass.  To facilitate subsequent scientific applications of the database, we also compile a number of ancillary physical properties of the galaxies, including their optical morphology, stellar mass, and various diagnostics of star formation activity.  We use the homogeneous catalog of \HI\ parameters to examine the statistical properties of profile asymmetry and shape.  Across the full sample, which covers a wide range of stellar masses and environments, statistically significant \HI\ profile asymmetry is detected in $\sim 20\%$ of the galaxy population.  The global \HI\ profiles are $35.2 \pm 0.3\%$ single-peaked, $26.9 \pm 0.3\%$ flat-topped, and $37.9 \pm 0.3\%$ double-horned.  At a given inclination angle, double-horned profiles are preferentially associated with galaxies of higher stellar mass or optical concentration, while galaxies of lower mass or concentration tend to have single-peaked profiles. 
\end{abstract}

\section{Introduction} 
\label{sec:intro}

The integrated 21~cm \HI\ spectrum of a galaxy encodes valuable information about its internal and external properties.  Apart from basic data on radial velocity, neutral atomic hydrogen gas mass, and rotation velocity, which are critical for many astrophysical and cosmological applications (e.g., \citealt{TullyFisher1977, Haynes2018ApJ...861...49H, Koribalski2004AJ....128...16K, Meyer2004MNRAS.350.1195M, Springob2005ApJS..160..149S, Wang2020ApJ...890...63W}), the detailed shape and asymmetry of the line profile also serve as effective tracers of galaxy kinematics and dynamics.

The extended and fragile \HI\ disk can be distorted easily by external perturbations, such as tidal interactions (\citealt{Hess2017MNRAS.464..957H}; \citealt{Sorgho2017MNRAS.464..530S}; \citealt{Bok2019MNRAS.484..582B}), ram pressure stripping (\citealt{Scott2010MNRAS.403.1175S}; \citealt{Kenney2015AJ....150...59K}), and gas accretion (\citealt{Bournaud2005A&A...438..507B}). The gas distribution also responds internally to feedback from stars (\citealt{Ashley2017AJ....153..132A}) and active galactic nuclei (\citealt{Morganti2017FrASS...4...42M}), and the non-axisymmetric gravitational potential from spiral arms \citep{Laine1998MNRAS.297.1041L} and bars \citep{Saha2007MNRAS.382..419S, Newnham2020MNRAS.492.4697N}. 

The asymmetry of the integrated \HI\ profile offers a blunt yet convenient tool to quantify the degree of dynamical perturbation on the galactic \HI\ disk (e.g., \citealt{Richter1994A&A...290L...9R, Haynes1998AJ....115...62H, Andersen2009ApJ...700.1626A}). It has been widely used historically. Measurement techniques vary greatly, from simple visual inspection \citep{Richter1994A&A...290L...9R} to quantitative prescriptions based on flux (e.g., \citealt{Haynes1998AJ....115...62H, Matthews1998AJ....116.1169M, Bournaud2005A&A...438..507B, Espada2011A&A...532A.117E}), line width (e.g., \citealt{Andersen2009ApJ...700.1626A}), and central velocity (e.g., \citealt{Haynes1998AJ....115...62H, Deg2020MNRAS.495.1984D}). The various asymmetry parameters typically are correlated with each other \citep{Haynes1998AJ....115...62H, Andersen2009ApJ...700.1626A}. 

Several complications affect the interpretation of previous investigations of \HI\ asymmetry. Besides the difficulty of comparing measurements made using disparate asymmetry parameters, there is also no commonly adopted quantitative criteria for the level of asymmetry that is deemed significant. Systematic biases due to signal-to-noise ratio (S/N) may be important, but, with a few exceptions (e.g., \citealt{Watts2020MNRAS.492.3672W}), these biases are usually neglected.  Sample selection also matters, for the \HI\ properties of galaxies vary strongly across the diverse galaxy population and the wide range of environments that they inhabit \citep{SaintongeARAA}.  These factors present a challenge to assessing the prevalence of \HI\ asymmetry in the galaxy population.  For instance, early studies have reported an asymmetry fraction of more than 50\% among isolated galaxies (e.g., \citealt{Richter1994A&A...290L...9R, Haynes1998AJ....115...62H, Matthews1998AJ....116.1169M}), which supports the notion that a galaxy's asymmetric gas distribution is long-lived (\citealt{Baldwin1980MNRAS.193..313B, Richter1994A&A...290L...9R}).  However, other studies find much lower asymmetry fractions ($\sim 10\%-30\%$; \citealt{Bournaud2005A&A...438..507B, Espada2011A&A...532A.117E, Bok2019MNRAS.484..582B, Watts2020MNRAS.492.3672W}), even in cluster environments (16\%--26\%; \citealt{Scott2018MNRAS.475.4648S}) and among galaxy pairs (13\%--27\%; \citealt{Bok2019MNRAS.484..582B, Zuo2022}).  The lack of consensus on the observational results has sparked debate on the physical drivers of profile asymmetry and their dependence on environment (\citealt{Espada2011A&A...532A.117E, Reynolds2020MNRAS.499.3233R, Watts2020MNRAS.492.3672W}).  

In contrast to asymmetry, for which an extensive body of work exists, relatively less attention has been devoted to characterizing the shape of the \HI\ line profile, or to discussion of its significance.  It is readily apparent that the integrated profile of the 21~cm line generally falls into one of three generic types---single-peaked, flat-topped, and double-horned---and various schemes have been developed to quantify them (e.g., \citealt{Shostak1977A&A....58L..31S, Stewart2014A&A...567A..61S, WestmeierBF2014MNRAS.438.1176W, El-Badry2018MNRAS.477.1536E, Dutton2019MNRAS.482.5606D, Yu2020ApJ...898..102Y}).  With few exceptions (e.g., \citealt{El-Badry2018MNRAS.477.1536E}), however, the diagnostic potential contained in the profile shape has been rarely exploited.  To first order, the shape of the \HI\ profile is determined by the inclination angle along the line-of-sight and the velocity and spatial distribution of the line-emitting gas.  Even in the absence of any spatial resolution, \cite{Yu2022} demonstrate that useful statistical inferences can be placed on the spatial distribution of the \HI\ using a sufficiently large sample of galaxies once certain global properties can be constrained.

In the last two decades, a number of single-dish surveys have secured global \HI\ spectra for tens of thousands of galaxies, chief among them the \HI\, Parkes All-Sky Survey (HIPASS: \citealt{Koribalski2004AJ....128...16K}; \citealt{Meyer2004MNRAS.350.1195M}; \citealt{Wong2006MNRAS.371.1855W}), the GALEX Arecibo SDSS Survey (xGASS: \citealt{Catinella2008AIPC.1035..252C, Catinella2012A&A...544A..65C, Catinella2013MNRAS.436...34C, Catinella2018MNRAS.476..875C}), and the Arecibo Legacy Fast ALFA survey (ALFALFA: \citealt{Haynes2011AJ....142..170H, Haynes2018ApJ...861...49H}). With over 31,000 detections, the ALFALFA survey is the largest blind \HI\, survey to date. Although the sample is biased in favor of star-forming galaxies (e.g., \citealt{Huang2012ApJ...756..113H, Zhang2019ApJ...884L..52Z}), which are more \HI-rich than typical nearby galaxies \citep{Catinella2010MNRAS.403..683C, Catinella2018MNRAS.476..875C} or galaxies in the Virgo cluster (\citealt{Cortese2011MNRAS.415.1797C}), ALFALFA nonetheless offers an opportunity to revisit the aforementioned issues, which have previously been handicapped by small-number statistics.  A homogeneous, systematic analysis of the ALFALFA database paves the way for even more ambitious exploration using upcoming surveys by the Square Kilometre Array (SKA: \citealt{Dewdney2009IEEEP..97.1482D}) and its pathfinders (e.g., ASKAP: \citealt{Johnston2007PASA...24..174J}, WALLABY: \citealt{Koribalski2020Ap&SS.365..118K}), the Five-hundred-meter Aperture Spherical radio Telescope (FAST: \citealt{Nan2011IJMPD..20..989N}), and Apertif \citep{Adams2018AAS...23135404A}.

This work utilizes the curve-of-growth (CoG) method developed by \citet{Yu2020ApJ...898..102Y} to perform a uniform analysis of the \HI\ spectra of all galaxies reliably detected by ALFALFA that also have secure optical counterparts.  Our primary aim is to extract robust parameters to quantify the line asymmetry and the profile shape, paying special attention to estimating and correcting for systematic uncertainties.  For the subset of galaxies that overlap with the Sloan Digital Sky Survey (SDSS: \citealt{York2000AJ....120.1579Y}), we take advantage of the ancillary optical information to derive useful physical parameters such as rotation velocities and dynamical masses, as well as compiling stellar masses (\mstar), star formation rates (SFRs), and other stellar population diagnostics that will serve as the foundation for forthcoming scientific applications of this database.

We describe our sample construction in Section~\ref{sec:sample}.  Section~3 presents the spectral measurements, error analysis, corrections for systematic uncertainties, and comparison with the literature. Statistical results on profile asymmetry and shape are highlighted in Section~4. Section~5 summarizes our main conclusions.

\section{Sample}
\label{sec:sample}

\subsection{The ALFALFA Survey}
\label{subsec:alfalfa}

The ALFALFA survey is the largest and most sensitive blind \HI\, survey to date, having detected 31,502 nearby objects with $z<0.06$.
The ALFALFA survey used the seven-horn Arecibo L-band Feed Array (ALFA) and the drift-scan mode to map $\sim 7000$ deg$^2$ with a beam size of 3$\farcm 3\times 3\farcm 8$. The survey covered two sky areas at high Galactic latitude, one in the northern and another in the southern Galactic hemisphere\footnote{The survey areas are $07^h30^m < {\rm R.A.} < 16^h30^m, \, \,  0^{\circ} < {\rm Decl.} < +36^{\circ}$ and $22^h < {\rm R.A.} < 03^h, \, \, 0^{\circ} < {\rm Decl.} < +36^{\circ}$, where R.A. and Decl. are the equatorial coordinates (J2000).}. The survey blindly searched for line emission within the frequency range $1335-1435$ MHz, which corresponds to the heliocentric velocities $-2000$ \kms\ $< cz < 18,000$ \kms. The on-source integration time of $\sim 48$~s per beam solid angle reached a noise level of 1.6 mJy.  With a channel width ($\delta V$) of 5.5 \kms, the final smoothed velocity resolution is 10 \kms, after Hanning smoothing.  The survey had three data releases, covering 40\%, 70\%, and 100\% of the final survey area, respectively \citep{Haynes2011AJ....142..170H, Haynes2018ApJ...861...49H}.  
%This study uses the spectra from the final data release, which consists of 31,501 sources with $z< 0.06$
%YY
This study uses 31,501 spectra from the final data release\footnote{We omit AGC~715637 because its spectrum is in pure absorption.}.

Except for 344 optically dark objects, the ALFALFA team identified the optical counterpart for each \HI\, detection by considering its position, optical color, and morphology from available catalogs and databases, such as the SDSS, SkyView (\citealt{McGlynn1994ASPC...61...34M}), the Second Palomar Observatory Digital Sky Survey (POSS-II; \citealt{Djorgovski1998wfsc.conf...89D}), and the NASA Extragalactic Database (NED)\footnote{{\url https://ned.ipac.caltech.edu}}. The accuracy of optical positions assigned by ALFALFA are normally 3$^{\prime \prime}$ or better \citep{Haynes2011AJ....142..170H}.  The survey contains 25,433 sources of quality code 1 and 6068 of quality code 2.  Code 1 refers to detections with the highest quality, which is assessed by (a) consistent signal characteristics between the two independent polarizations, (b) a well-defined spatial extent that is consistent with or larger than the telescope beam, (c) spectral profile relatively free from radio frequency interference (RFI), and (d) a minimum S/N of $\sim 6.5$, which is required for a detection reliability of 95\% \citep{Saintonge2007AJ....133.2087S}. No constraints are placed on having a matching optical redshift. Code 2 includes objects with ${\rm S/N} \lesssim 6.5$, but it must have an optical counterpart with a known optical redshift coincident with that measured in \HI. 

The ALFALFA catalog \citep{Haynes2018ApJ...861...49H} tabulates measurements of the total line flux, central velocity ($V_{c\alpha}$), and line widths at 20\% ($W_{20}$) and 50\% ($W_{50}$) of the line peak. For each source, the minimum and maximum line widths measured by the ALFALFA team are denoted by $W_{\rm min}$ = min($W_{20}$, $W_{50}$) and $W_{\rm max}$ = max($W_{20}$, $W_{50}$), respectively. For galaxies with $V_{c\alpha} \geq 6000$ \kms, galaxy distances are computed from $cz_{\rm CMB}/H_0$, where $cz_{\rm CMB}$ is the recessional velocity measured in the cosmic microwave background reference frame \citep{Lineweaver1996ApJ...470...38L} and $H_0 = 70$ \kms\ Mpc$^{-1}$ is the Hubble constant. If $V_{c\alpha} < 6000$ \kms, distances derive from literature analysis of the peculiar motions of galaxies, groups, and clusters (\citealt{Masters2005PhDT.........2M, Mei2007ApJ...655..144M, Springob2007ApJS..172..599S,  Hallenbeck2012AJ....144...87H}) and the Tully–Fisher relation \citep{Tully2013AJ....146...86T}.

\begin{figure}
\epsscale{1.1}
\plotone{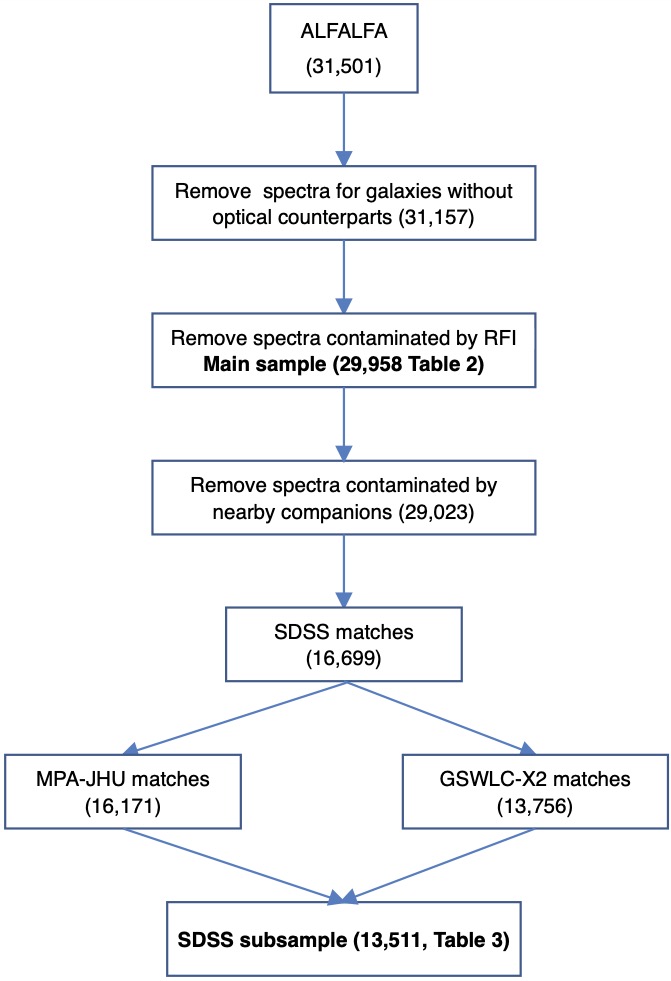}
\caption{Flowchart of sample selection process.}
\label{fig:flow}
\end{figure}

\subsection{The Main Sample}
\label{subsec:prim}

Figure~\ref{fig:flow} gives a flowchart of our process of sample construction.  Starting from the original ALFALFA catalog of 31,501 galaxies, we applied the following additional selection steps to derive a main sample of 29,958 galaxies.

\begin{itemize}

\item{We remove 344 objects without optical counterparts.  From \citet{Haynes2011AJ....142..170H}, these objects typically have code 1 and flag ``U'', which means that they have ${\rm S/N} \gtrsim 6.5$ and lie within the SDSS DR7 footprint but do not have an identified optical counterpart. Most of these objects are probably tidal dwarfs, low-surface brightness dwarfs \citep{Jones2019A&A...632A..78J}, or tidal debris \citep{Durbala2020AJ....160..271D}. An object that belongs to this category AGC~229101, which \cite{Leisman2021AJ....162..274L} associated with a faint, blue galaxy with an \HI-to-stellar mass ratio of $\sim 100$, possibly formed through tidal interaction with neighboring objects or merged from two dark \HI\, clouds. Due to the unknown nature and optical properties of these objects, we exclude them from subsequent analysis.}

\item{We remove 1199 galaxies strongly contaminated by RFI. We use the spectral weight to evaluate the level of RFI contamination in each channel.  The minimum acceptable spectral weight is 0.5 \citep{Haynes2018ApJ...861...49H}, and a spectral weight larger than 0.9 \citep{Giovanelli2007AJ....133.2569G} implies that $\sim 85\%$ of the total band pass is free of RFI.  We reject spectra having bad channels with velocities within $V_{c\alpha}\pm W_{\rm min}$/2, or within [$V_{c\alpha}-W_{\rm max}$, $V_{c\alpha}-W_{\rm min}$/2] or [$V_{c\alpha}+W_{\rm min}$/2, $V_{c\alpha}+W_{\rm max}$] and the fraction of bad channels is more than 30\% or the standard deviation of bad channels exceeds twice the spectral noise level. Spectra affected by a low fraction of outlying bad channels are kept but flagged by ``r''. }

\end{itemize}

Within the main sample, we take special care to identify 935 galaxies whose \HI\, spectra are potentially blended by that of nearby companions.  The velocity difference of two galaxies in merger pairs is smaller than 500 \kms\ (e.g., \citealt{Patton2000ApJ...536..153P, Lambas2003MNRAS.346.1189L, Ellison2008AJ....135.1877E, Ellison2010MNRAS.407.1514E, Bok2019MNRAS.484..582B}).  We consider nearby neighbors as candidate physical companions if it has an optical position separation smaller than the Arecibo beam size of 3$\farcm$5 and velocity difference less than 500 \kms, using the optical positions and recession velocities from the ALFALFA catalog. If the central velocity difference is smaller than the sum of $W_{\rm min}/2$ of the target galaxy and its apparent companion, the apparent companion is considered a candidate companion.  If the central velocity difference is smaller than the sum of $3W_{\rm max}/4$ of the target galaxy and its apparent companion, and the median value of the flux intensities in the likely overlapping channels, which lie between the two velocity centers but beyond the velocity center range $\pm W_{\rm min}$/2 of each, is larger than the spectral noise level from the ALFALFA catalog, then the spectrum is considered to be affected by an apparent companion. In the above two situations, the \HI\ spectra of galaxies are considered as contaminated by nearby companions.  These objects, flagged by ``c'', are retained in the main sample, to facilitate subsequent analysis that may or may not wish to include them, depending on the scientific goals. 
Galaxies having a projected neighbor within the Arecibo beam but whose \HI\ signal does not overlap with that of the target galaxy are flagged by ``p''.

\begin{figure*}[ht!]
\centering
\includegraphics[width=1.0\textwidth]{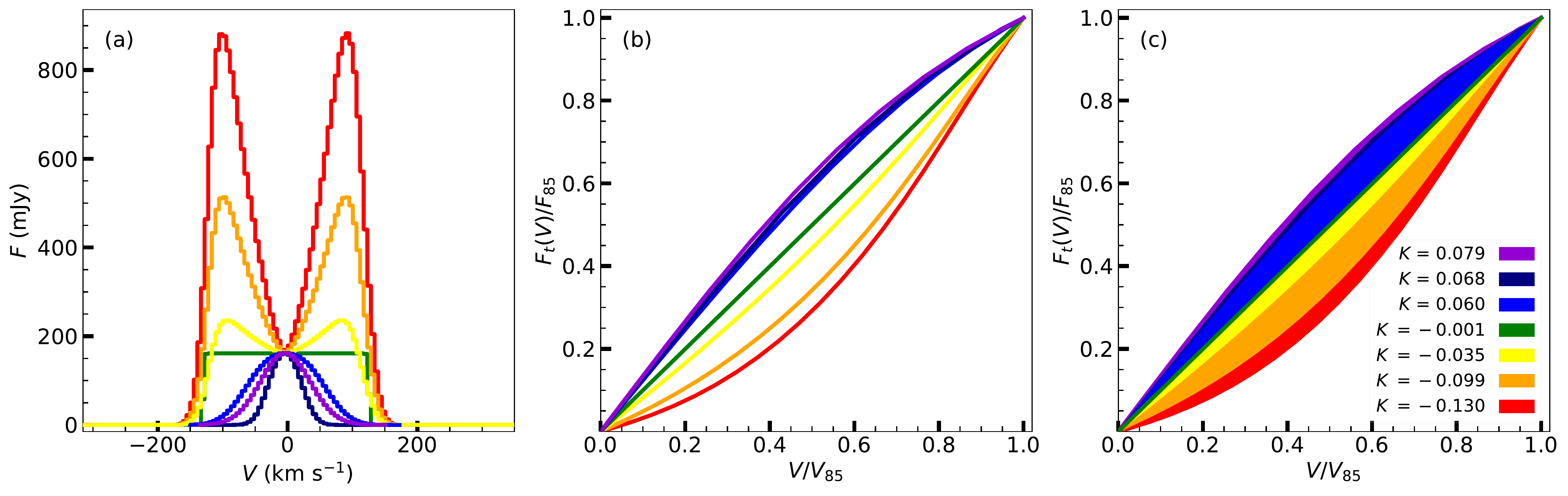}
\caption{Illustration of the method to measure the profile shape parameter $K$, for (a) seven mock profiles that vary from single-peaked to flat-topped and double-horned and (b) their corresponding normalized CoGs. In panel (c), the shaded areas are the values of $K$, which are given in the lower-right corner of the panel; areas below the diagonal line are negative and correspond to double-horned profiles, while those above are positive and represent single-peaked profiles.}
\label{fig:kappaS}
\end{figure*}

\subsection{The SDSS Subsample}
\label{subsec:cross} 

A significant portion of ALFALFA overlaps, by design, with SDSS.  This enables a variety of science applications. To exclude the effects of the confusion and contamination, we remove galaxies whose \HI\ spectra are contaminated by nearby companions (flagged by ``c'') from the main sample. We generate an SDSS-matched subsample of galaxies based on projected separation and radial velocity.  We use a maximum projected optical position separation of 3$^{\prime\prime}$, the typical positional accuracy of the optical counterparts in ALFALFA \citep{Haynes2018ApJ...861...49H}, which can be easily satisfied by the astrometric accuracy of SDSS ($\sim 0\farcs1$ at $r < 22$ mag; \citealt{Pier2003AJ....125.1559P}).  We further require that there be spectroscopy available from SDSS DR16 (\citealt{AhumadaSDSSDR162020ApJS..249....3A}), and that the maximum velocity difference between the \HI\ and the optical not exceed 500 \kms, which is within the range of observed velocity offsets in bound galaxy pairs and galaxy mergers (e.g., \citealt{Patton2000ApJ...536..153P, Ellison2010MNRAS.407.1514E}).  The optical velocities from SDSS are accurate to $\sim$ 30--70 \kms\, (\citealt{York2000AJ....120.1579Y, AhumadaSDSSDR162020ApJS..249....3A}). 

We collect from SDSS DR16 basic optical properties of the galaxies, including $R_{90}$ and $R_{50}$, the radii enclosing 90\% and 50\% of the $r$-band Petrosian flux, and axis ratio $q$ in the $r$ band. The optical concentration index, $C\equiv R_{90}/R_{50}$, describes the optical light concentration. To quantify the strength of star formation in the central region of the galaxy, we obtain from the MPA-JHU database (\citealt{Kauffmann2003MNRAS.341...33K, Brinchmann2004MNRAS.351.1151B}) information on EW(H$\alpha$), the H$\alpha$ equivalent width, which probes the average specific SFR over the past 100 Myr (\citealt{Fumagalli2012ApJ...757L..22F,Marmol-Queralto2016MNRAS.460.3587M, Khostovan2021MNRAS.503.5115K}), $D_{n}4000$, the ratio of the average flux density at 4000--4100 \AA\ to that at 3850--3950 \AA\, \citep{Balogh1999ApJ...527...54B}, which is sensitive to the mean stellar age over the past 1 Gyr \citep{Kauffmann2003MNRAS.341...33K}, and the SFR within the 3$^{\prime\prime}$-diameter SDSS fiber (SFR$_{\rm in}$)\footnote{At a median distance of 137 Mpc for our sample, 3\arcsec\ corresponds to $\sim 2$ kpc.}, which is based on the extinction-corrected H$\alpha$ line emission and $D_{n}4000$.  We supplement the MPA-JHU measurements with total stellar masses and SFRs from the second version of the GALEX-SDSS-WISE Legacy Catalog with the deepest photometry (GSWLC-X2) of \cite{Salim2018ApJ...859...11S}, who fitted the spectral energy distribution of the global photometry from the mid-infrared, optical, and ultraviolet bands. The typical uncertainties of the stellar masses and SFRs from \citet{Salim2018ApJ...859...11S} are 0.042 dex and 0.064 dex, respectively.  All distance-dependent quantities are scaled to the distances adopted from ALFALFA.  The SFRs and stellar masses assume the stellar initial mass function of \cite{Chabrier2003PASP..115..763C}.  The final SDSS-matched subsample of 13,511 galaxies is summarized in Table~\ref{tab:opt}.

\begin{figure*}
\centering
\includegraphics[width=0.8\textwidth]{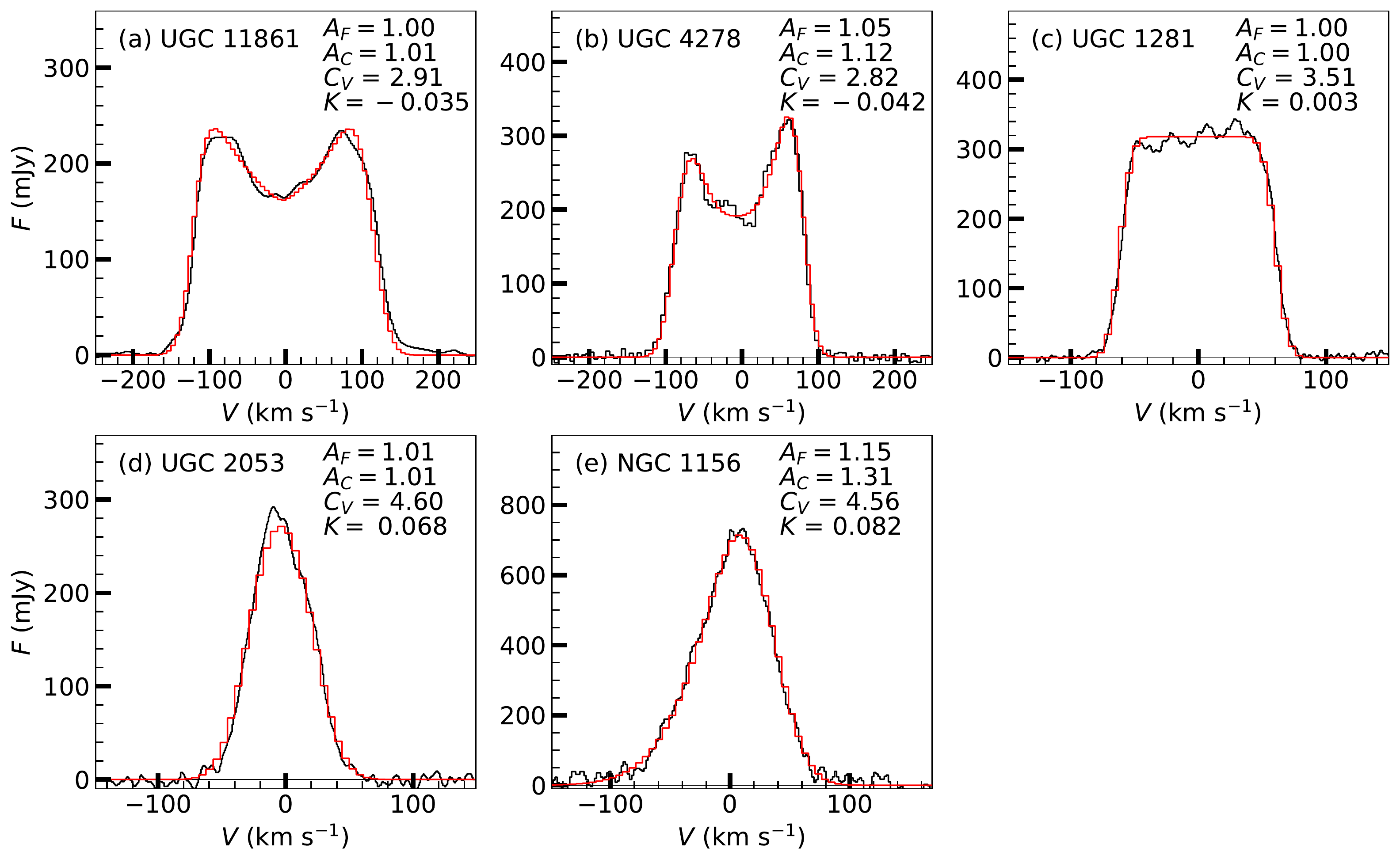}
\caption{Illustration of five typical profile shapes, as observed (black) and fit (red) with the busy function \citep{WestmeierBF2014MNRAS.438.1176W}. The profiles vary from (a) symmetric double-horned (UGC~11861) to (b) asymmetric double-horned (UGC~4278), (c) flat-topped (UGC~1281), (d) symmetric single-peaked (UGC~2053), and (e) asymmetric single-peaked (NGC~1156). The values of asymmetry and shape parameters for the mock spectra are shown in the upper-right corner.}
\label{fig:typical}
\end{figure*}

\section{\HI\ Profile Measurements}
\label{sec:measurements}

\subsection{Curve-of-Growth Analysis}
\label{subsec:cog}

Our analysis uses the CoG method of \citet{Yu2020ApJ...898..102Y}, which integrates the flux intensity as a function of velocity from the line center outward to both the blue and red sides of the profile, to measure several non-parametric quantities to describe the \HI\ emission line.  This method is close in spirit to those of \cite{Courteau1997AJ....114.2402C} and \cite{Andersen2009ApJ...700.1626A}.  We closely follow the formalism of \citet{Yu2020ApJ...898..102Y}, except that no additional subtraction of the baseline is necessary because the baselines of the ALFALFA spectra are already relatively clean and flat, and our quality check based on spectral weight (Section~\ref{subsec:prim}) has successfully mitigated against RFI contamination. We mirror the negative side of flux intensity distribution to the positive side and fit the resulting distribution with a Gaussian function, whose best-fit standard deviation gives an estimate of the noise level ($\sigma_{\rm spec}$) of the spectra.

Some slight modifications are made to the automated procedure to delineate the limits of the CoG integration, optimized for and tested using ALFALFA data.  We search for emission signal within a range of $\pm 500$ \kms\ around the central velocity given in the ALFALFA catalog.  We first select segments of at least three consecutive channels having positive flux intensities.  Designating the segment with the highest integrated flux as reference, only segments with integrated flux or mean integrated flux larger than 0.5 times the value of the reference segment are considered as candidate emission signal.  Starting from the central segment, adjacent segments with separations less than 50 \kms\ are merged in succession, until all segments have been combined into a single final segment spanning $N$ channels.  To facilitate subsequent calculations, we pad the final segment with $0.5\, N$ channels on both ends, resulting in a final data array of $2\, N$ channels.

\begin{figure*}
\centering
\includegraphics[width=1.0\textwidth]{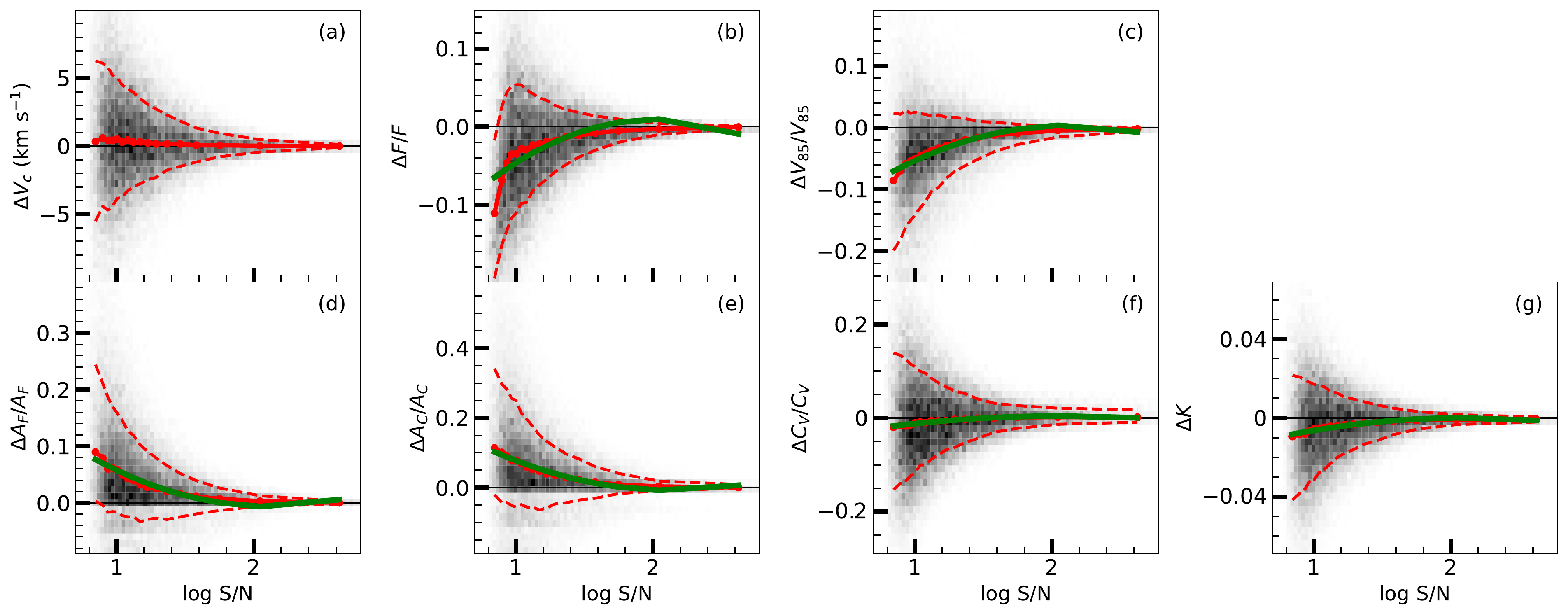}
\caption{Simulations using mock spectra to test the sensitivity of our parameter measurements to variations in S/N. The mock spectra were generated with five types of representative profiles (flat-topped, symmetric single-peaked, asymmetric single-peaked, symmetric double-horned, and asymmetric double-horned). The panels plot (a) absolute deviations for central velocity $V_c$ and fractional deviations for (b) flux $F$, (c) velocity width $V_{85}$, (d) flux asymmetry $A_F$, (e) flux distribution asymmetry $A_C$, (f) profile concentration $C_V$, and absolute deviation for (g) profile shape $K$. The black horizontal line denotes zero offset. The red solid lines show the median, and the dashed lines show the 16th and 84th percentiles of the distribution. Green curves in panels (b)–-(g) are polynomial fits to the median distribution of total flux, line widths, asymmetry, and profile shape parameters (Equation 1).}
\label{fig:test}
\end{figure*}

With the limits of integration thus defined, we calculate the flux intensity-weighted central velocity $V_c$ and generate the CoG\footnote{Roughly 4\% of the spectra, typically on account of their low S/N or relatively broad emission, cannot be measured automatically by the CoG method. For these objects we set the velocity ranges manually.}.  The total line flux $F$ follows from the median value of the flat part of the CoG.  Line widths are specified as the velocity width that encloses characteristic percentages of the total flux. This catalog provides only $V_{85}$, the velocity width that captures 85\% of the total flux, which most effectively represents the rotation velocity of the galaxy \citep{Yu2020ApJ...898..102Y}.  We quantify the asymmetry of the line using the red and blue sides of the CoG.  The asymmetry parameter $A_F$ is defined as the larger of the ratios of the integrated flux of the one side of the line relative to the other, while $A_C$ is the larger of the ratios of the slopes of the rising part of the CoG of the two sides.  By definition, $A_F \geq 1$ and $A_C \geq 1$. These two asymmetry parameters are broadly equivalent to previous intensity-weighted methods (e.g., \citealt{Giese2016MNRAS.461.1656G, Deg2020MNRAS.495.1984D}). The parameter $C_V$ is the ratio of line widths $V_{85}$ and $V_{25}$, where $V_{25}$ is  the line width enclosing 25\% of the total flux of the CoG. Thus, $C_V$ characterizes the degree of concentration of the line profile: $C_V = 85/25 = 3.4$ for a flat-topped (perfectly rectangular) profile.  Being non-parametric, $C_V$ affords a more stable and less model-dependent description of the profile shape than other choices (e.g., \citealt{Stewart2014A&A...567A..61S, WestmeierBF2014MNRAS.438.1176W}).  

Here we introduce another new, non-parametric measure of profile shape, one that can discern even more subtle features of the line profile than $C_V$.  Normalizing the CoG by $V_{85}$ and the integrated flux to 85\% of the total flux, we define the profile shape $K$ as the integrated area between the normalized CoG and the diagonal line of unity in Figure~\ref{fig:kappaS}b; $K$ is positive or negative if the normalized CoG is above or below the line of unity.  To illustrate how the profile shape influences $K$, we simulate seven typical \HI\, profiles (Figure~\ref{fig:kappaS}a) using the ``busy function'' (\citealt{WestmeierBF2014MNRAS.438.1176W}) and show their normalized CoGs in Figure~\ref{fig:kappaS}b.  As can be seen from Figure~\ref{fig:kappaS}c, a double-horned profile yields  $K<0$, a perfectly flat-topped profile gives $K=0$, while a single-peaked line results in $K>0$. With increasing $K$, the line shape transitions from a broad, double-horned profile to a narrow, single-peaked profile.

The parameters $C_V$ and $K$ quantify the profile shape: larger values correspond to larger profile concentration. The \HI\, profile shape is, to the first order, governed by the inclination angle of the disk and the velocity and spatial distribution of the gas \citep{El-Badry2018MNRAS.477.1536E}. In a companion paper, \cite{Yu2022} use the present catalog to investigate the connection between \HI\ distribution and star formation activity in nearby galaxies.

\begin{deluxetable}{cccc}
\tablenum{1}
\tabletypesize{\footnotesize}
\tablecolumns{4}
\tablecaption{Measurement Uncertainties}
\tablehead{
\colhead{Parameter} &
\colhead{Systematic} &
\colhead{Systematic} &
\colhead{Statistical} 
\\
\colhead{} &
\colhead{($5< {\rm S/N} < 10$)} &
\colhead{(${\rm S/N} \geq 10$)} &
\colhead{} 
\\
\colhead{(1)} &
\colhead{(2)} &
\colhead{(3)} &
\colhead{(4)} 
}
\startdata
$V_c$      &  6.0  & 3.0  & 2.8 \\
$F$   	   &  10\% & 5\%  & 4\% \\
$V_{85}$   &  12\% & 6\%  & 3\% \\
$A_F$      &  12\% & 7\%  & 4\% \\
$A_C$      &  20\% & 10\% & 6\% \\
$C_V$      &  14\% & 7\%  & 5\% \\
$K$   &  0.029&0.015 & 0.011 \\
\enddata
\tablecomments{Col. (1): Measured parameter. The uncertainties for $V_c$ and $K$ are absolute, while those for the other parameters are fractional.  Col. (2): Typical systematic uncertainties due to S/N if $5< {\rm S/N} < 10$.  Col. (3): Typical systematic uncertainties due to S/N if ${\rm S/N} \geq 10$.  Col. (4): Typical statistical uncertainties derived from Monte Carlo simulations.}
\label{tab:uncert}
\end{deluxetable}

\begin{figure*}[t]
\centering
\epsscale{1.0}
\plotone{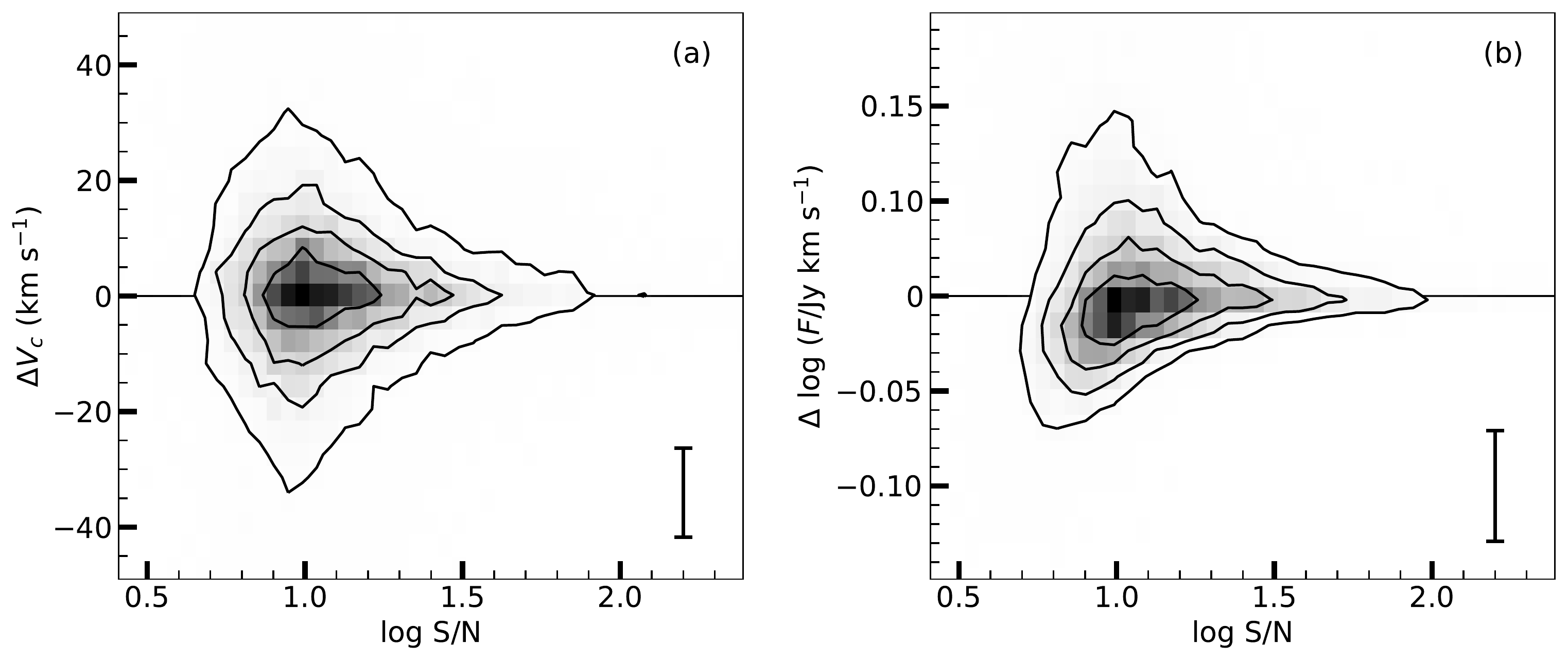}
\caption{Comparison between our measurements of (a) central velocity $V_c$ and (b) total flux $F$ with measurements from the ALFALFA catalog, as a function of the S/N derived by the CoG method. Representative error bars are given in the bottom-right corner of each panel. The contours in each panel show the levels of sample inclusion, which vary from 20\%, 40\%, 60\%, and 80\% from inside out.}
\label{fig:compALFA}
\end{figure*}

\subsection{Uncertainty Estimation and Systematic Corrections}
\label{subsec:uncertainties}

The parameters derived in this paper are affected by both statistical and systematic uncertainties. The statistical uncertainties for the parameters of each spectrum are estimated from a set of 50 mock spectra by adding to the original spectrum random noise following a Gaussian distribution with standard deviation equivalent to the noise level $\sigma_{\rm spec}$.  The mock spectra are analyzed in the same way as the real spectra, and the standard deviation of the distribution of each parameter is taken as its statistical uncertainty.

Several sources of systematic uncertainty need to be considered. We generate mock spectra to quantify the systematic effect of spectral noise on our parameter measurements. As in \cite{Yu2020ApJ...898..102Y}, we use the busy function \citep{WestmeierBF2014MNRAS.438.1176W} to fit the observed spectra of five galaxies whose \HI\ spectra span a diversity of observed profile shapes, from flat-topped to symmetric single-peaked, asymmetric single-peaked, symmetric double-horned, and asymmetric double-horned. The observed \HI\ spectra (NGC~1156: \citealt{1978ApJ...223..391D}; UGC~2053, UGC~1281, UGC~11861: \citealt{TifftCocke1988ApJS...67....1T}; UGC~4278: \citealt{Huchtmeier2005AA...435..459H}), along with their busy function fits, are shown in Figure~\ref{fig:typical}.  The derived values of profile asymmetry and shape approximately evenly cover more than 60\% of the parameter distributions of the entire survey (see Figures~\ref{fig:asydistribution} and \ref{fig:shapedistribution}). As with ALFALFA, the spectra have a channel width of 5.5 \kms.  We add noise and generate $10^4$ mock spectra with ${\rm S/N} \approx 5$ to 600. We derive CoG parameters for the mock spectra, and then compare the recovered parameters with the input values, as a function of S/N (Figure~\ref{fig:test}).

The measurements of central velocity show no systematic deviation, although the scatter becomes predictably larger with decreasing S/N. By contrast, both the total flux and line width are underestimated slightly ($\lesssim 10\%$) at low S/N, owing to the loss of the faint wings of the line profile in noisy spectra.  The asymmetry parameters $A_F$ and $A_C$ are systematically overestimated with decreasing S/N, consistent with the finding of \citet{Watts2020MNRAS.492.3672W} for a differently defined asymmetry parameter, but the systematic bias never exceeds 10\% for ${\rm S/N} \gtrsim 5$.  The profile shape parameters are systematically underestimated, although the effect is quite marginal ($\lesssim 2\%$ for $C_V$ and $\lesssim  0.01$ for $K$).  We also evaluate the systematic effects of adopting different criteria to detect segments for constructing the CoG (Section~\ref{subsec:cog}). In practice, we explore a range of choices of the threshold or mean flux ratio (relative to the brightest segment) and the velocity separation between two adjacent segments, perturbing their values around the fiducial default values.  We conclude that the uncertainties introduced by these criteria are negligible compared with other sources of uncertainty, and hence we neglect it from further consideration.  While our method of error analysis is identical to that in \citet{Yu2020ApJ...898..102Y}, the final systematic uncertainties are smaller in the current study because of improvements made to the signal selection algorithm for the CoG analysis (see Section~\ref{subsec:cog}). 

\begin{figure*}
\centering
\includegraphics[width=1.0\textwidth]{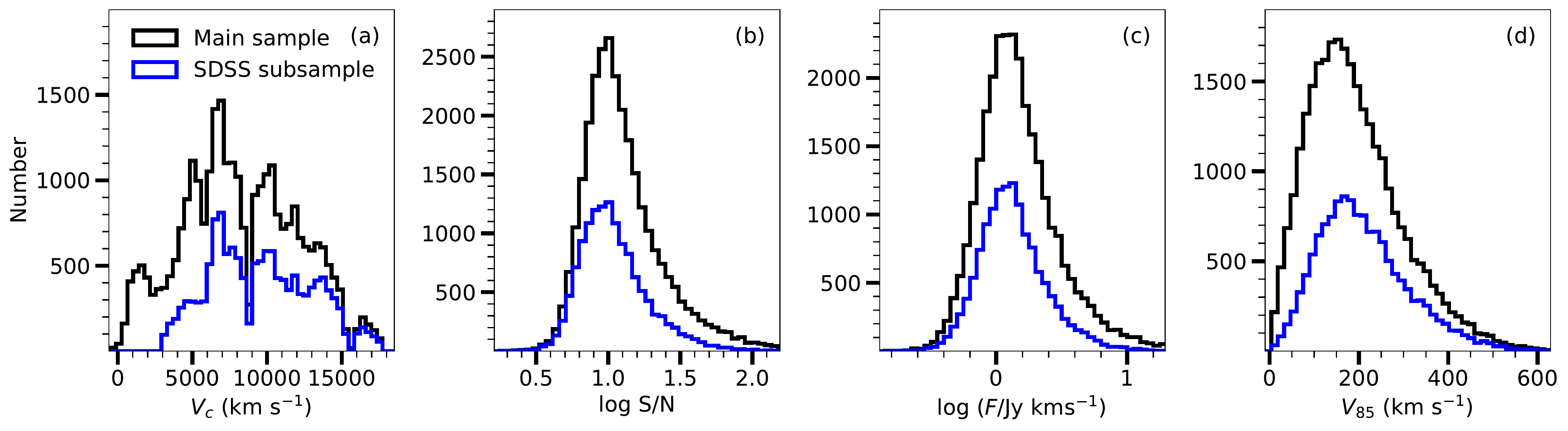}
\caption{Distribution of (a) central velocity $V_c$, (b) S/N, (c) total flux $F$, and (d) observed velocity width $V_{85}$ for the main sample (black) and SDSS subsample (blue).}
\label{fig:paraHist}
\end{figure*}

In light of the detectable and quantifiable systematic uncertainties identified for the measurements of total flux, line width, asymmetry, and profile shape, we correct for the systematic biases by fitting a second-order polynomial to the median trends obtained from the mock tests (green curves in Figure~\ref{fig:test}).  The corrections, applied only to spectra with ${\rm S/N}<40$ (the corrections are negligible for higher S/N), are as follows:

\begin{equation}
\begin{split}
\Delta F/F &= -0.053 (\log {\rm S/N})^2+ 0.216 \log {\rm S/N}- 0.210\\
\Delta V_{85}/V_{85} &= -0.045 (\log {\rm S/N})^2+ 0.193 \log {\rm S/N}- 0.201\\
\Delta A_F/A_F &= \,\,\,\,\,0.051 (\log {\rm S/N})^2-0.216 \log {\rm S/N}+0.223\\
\Delta A_C/A_C &= \,\,\,\,\, 0.065 (\log {\rm S/N})^2-0.281 \log {\rm S/N}+0.293\\
\Delta C_V/C_V &= -0.014 (\log {\rm S/N})^2+0.059 \log {\rm S/N}-0.057\\
\Delta K & = -0.005 (\log{\rm S/N})^2+0.022 \log{\rm S/N}-0.023.\\
\end{split}
\label{equ:corrections}
\end{equation}

\noindent
Unless otherwise noted, we correct the above parameters for these systematic deviations.

The final error budget of each parameter is the quadrature sum of the statistical uncertainties and systematic uncertainties, as summarized in Table~\ref{tab:uncert}. Following \citet{Springob2005ApJS..160..149S}, we include an additional uncertainty of 15\% for the total flux, to account for \HI\ self-absorption and calibration-related effects (beam attenuation, pointing, and flux calibration).

\subsection{Comparison with ALFALFA}
\label{subsec:correction}

We use the published ALFALFA catalog as an independent verification of our measurements.  This is possible only for the central velocity and the total flux. The other parameters are either new to our study (e.g., profile shape), not uniformly measured (e.g., asymmetry), or are defined differently (e.g., line widths).  Our central velocity $V_c$ is defined as the flux intensity-weighted central velocity, while the central velocity in the ALFALFA catalog is the midpoint of the channels with flux intensity equal to the 50\% level of each of the two peaks. Our total flux is derived by calculating the median integrated flux on the flat part of the CoG, which differs from the approach employed by ALFALFA.  Nevertheless, the two sets of measurements show no obvious systematic discrepancies with respect to each other (Figure~\ref{fig:compALFA}).  The scatter increases from $\Delta V_c = 0.4\pm 7.0$ \kms\, for ${\rm S/N}\geq 10$ to $\Delta V_c = 0.0\pm 10.8$ \kms\, for ${\rm S/N}<10$. As for the line flux, $\Delta \log F = 0.00\pm$0.01 dex for ${\rm S/N}\geq 10$, with the scatter and standard deviation increasing to $\Delta \log F=0.00\pm $0.03 dex for ${\rm S/N}<10$. The scatter is consistent with the typical uncertainties in Table~\ref{tab:uncert}.

\subsection{Final Measurements and Derived Physical Properties}

Figure~\ref{fig:paraHist} summarizes the basic properties of the main sample and SDSS subsample.  For the main sample, central velocities range from $V_c \approx 0$ to 18,000 \kms, which correspond to redshifts $z \approx 0-0.06$.  The spectra vary greatly in quality, from ${\rm S/N} \approx 4$ to more than 100, with integrated \HI\ fluxes spanning $F = 0.25$ to 10 Jy \kms.  The observed velocity widths cluster near $V_{85} \approx 200$ \kms, with some as low as 50 \kms\ and a long tail stretching out to $\sim 500$ \kms, which may be due to confusion from unrecognized nearby sources (e.g., \citealt{Brosch2011MNRAS.415..431B}).  The principal parameters measured for the main sample are given in Table~\ref{tab:co-add}.  

The SDSS subsample closely matches the main sample in its general characteristics (Figure~\ref{fig:paraHist}, blue histograms). The redshift range of the SDSS subsample is $z \approx 0.01-0.06$, which is mainly constrained by the matching with GSWLC-X2.  The availability of ancillary optical data allows us to deduce a number of useful physical properties for the galaxies (Table~\ref{tab:opt}). Of chief importance is the optically derived axis ratio, $q=b/a$, which can be used to estimate the inclination angle ($i$) of the galaxy. Following Hubble's (1926) prescription,

\begin{equation}
{\rm cos}^2 \,i = \frac{q^2-q^2_0}{1-q^2_0},
\label{equ:incl}
\end{equation}

\noindent
where $q_0$ is the intrinsic disk thickness. We estimate $q_0$ from the stellar mass following \citet{Sanchez-Janssen2010MNRAS.406L..65S}.  We assume that the \HI\ disk is coplanar with the stellar disk. This is generally a reasonably good assumption in most galaxies (e.g., \citealt{Verheijen2001A&A...370..765V}), even though in some early-type galaxies the stars and \HI\, can be severely misaligned (e.g., \citealt{Morganti2006MNRAS.371..157M}).  We correct the observed velocity width for the effects of projection, redshift, instrumental resolution, and turbulence broadening \citep{Yu2020ApJ...898..102Y}, 

\begin{equation}
V_{85c} = {{(V_{85} - V_{\rm inst})/(1+z) - V_{\rm turb}}\over{2 \ {\rm sin}\ i}},
\end{equation}

\noindent
where $V_{\rm inst}$ and $V_{\rm turb}$ are line-width correction terms for the instrumental resolution and turbulence motion, respectively. At a given instrumental resolution (10 \kms; \citealt{Haynes2018ApJ...861...49H}) and turbulence velocity dispersion (10 \kms; \citealt{Lelli2016AJ....152..157L}), the correction terms depend on the observed line width (see Figure~6 of \citealt{Yu2020ApJ...898..102Y}). 

With $V_{85c}$ in hand, we obtain the galaxy rotation velocity ($V_{\rm rot}$) from the empirical calibration outlined in \cite{Yu2020ApJ...898..102Y}.  Finally, we calculate the dynamical mass $M_{\rm dyn} = V_{\rm rot}^2R/G$, assuming, as in 
\cite{Yu2020ApJ...898..102Y}, that $R = R_{\rm HI}$, the characteristic \HI\, radius that obeys the \HI\ mass-size relation of \citet{Wang2016MNRAS.460.2143W}.

\begin{figure*}[!htbp]
\epsscale{1.}
\plotone{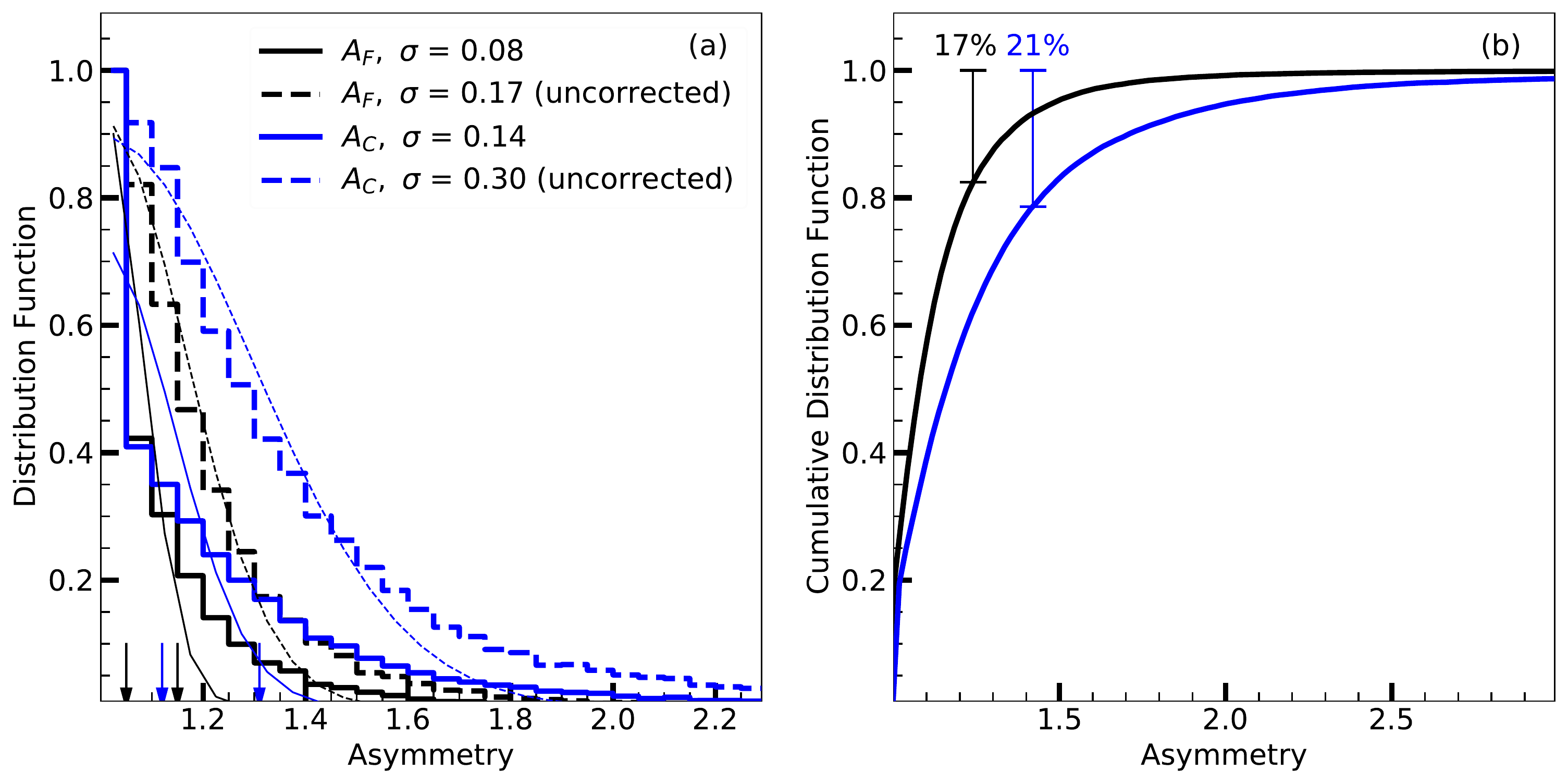}
\caption{(a) Distribution function and (b) cumulative distribution function of the asymmetry parameters $A_F$ (black) and $A_C$ (blue).  In panel (a), the standard deviation of the half-Gaussian fits of the distributions (thin curves) are given in the legends, and the black and blue arrows mark the values of $A_F$ and $A_C$ of the two asymmetry profiles in our simulations (Figures~3 and 4). The fraction of the sample that is asymmetric, defined as $A_F > 1.24$ and $A_C > 1.42$, is denoted in panel (b). }
\label{fig:asydistribution}
\end{figure*}

\begin{figure*}[!htbp]
\epsscale{0.8}
\plotone{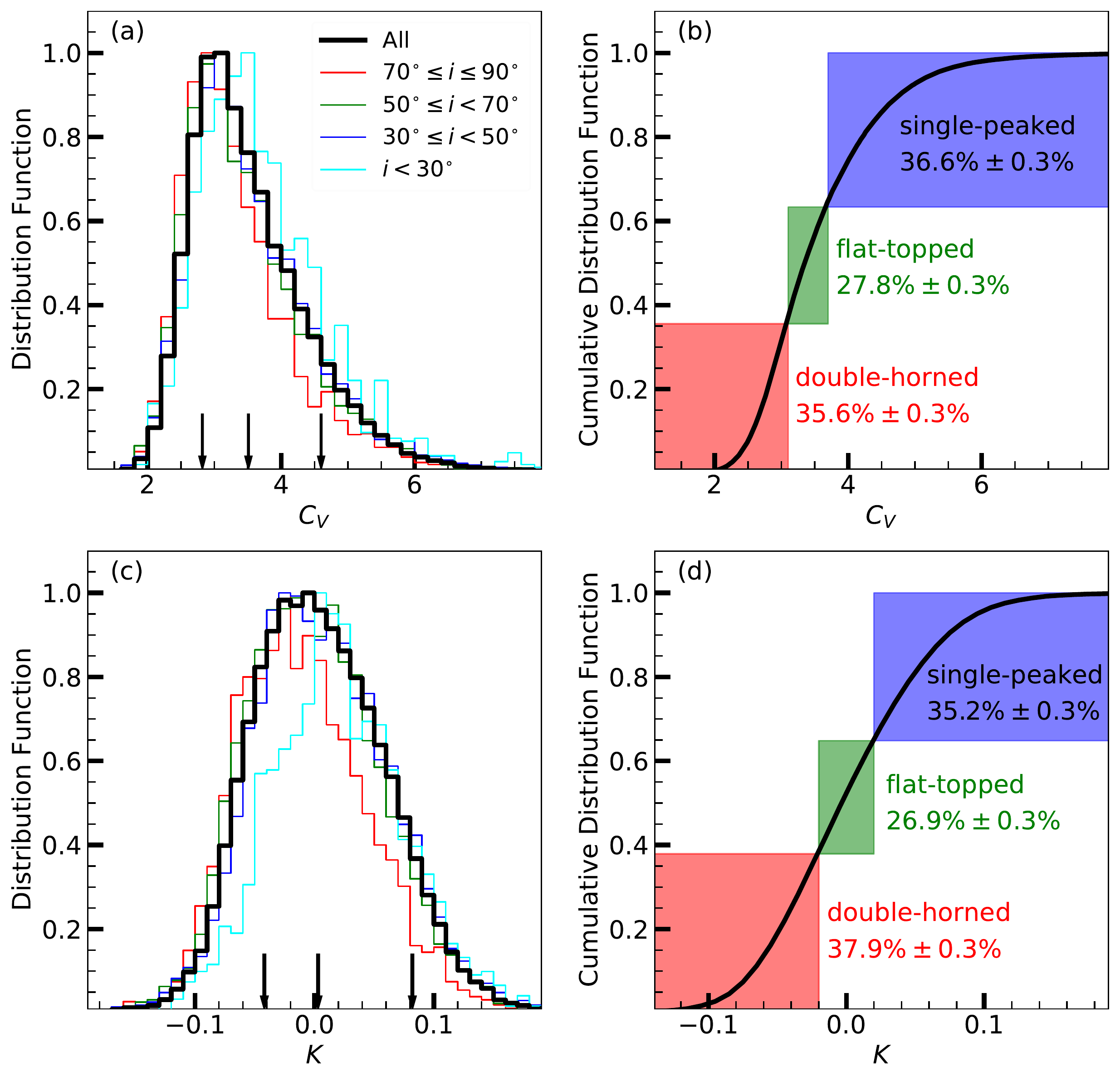}
\caption{Distribution function (left) and cumulative distribution function (right) of the profile shape parameters $C_V$ (top) and $K$ (bottom). For the distribution functions, objects of the main sample are plotted (thick black histogram), as well as for SDSS subsamples binned by inclination angle: $70^{\circ}\leq i\leq 90^{\circ}$ (red), $50^{\circ}\leq i< 70^{\circ}$ (green), $30^{\circ}\leq i< 50^{\circ}$ (blue), and $i< 30^{\circ}$ (cyan). In panels (a) and (c), the black arrows show the values of $C_V$ and $K$ of three different profile shapes used in our simulations (Figures~3 and 4).}
\label{fig:shapedistribution}
\end{figure*}

\section{Results}
\label{sec:sa}

\subsection{Statistical Properties of \HI\ Profile Asymmetry}

Asymmetric gas distribution or kinematics can be induced by environmental effects (e.g, \citealt{Cortese2021PASA...38...35C}, and references therein), major and minor mergers (e.g., \citealt{Jog1992ApJ...387..152J, Barnes2002MNRAS.333..481B, Boomsma2005A&A...431...65B, Robertson2006ApJ...645..986R}), gas accretion (\citealt{Bournaud2005A&A...438..507B, Sancisi2008A&ARv..15..189S}), internal perturbations by stellar feedback (\citealt{Ashley2017AJ....153..132A}), active galactic nuclei feedback (\citealt{Morganti2017FrASS...4...42M}), and non-axisymmetric perturbations to the gravitational potential (\citealt{Baldwin1980MNRAS.193..313B, Hayashi2006MNRAS.373.1117H}) due to a bar (\citealt{Saha2007MNRAS.382..419S, Newnham2020MNRAS.492.4697N}) or spiral arms (\citealt{Laine1998MNRAS.297.1041L}).  

Figure~\ref{fig:asydistribution}a shows the distribution of the asymmetry parameters $A_F$ and $A_C$ of the main sample, which is characterized by a dominant peak centered at 1, followed by a long, extended tail toward higher values.  We emphasize the importance of S/N correction (Section~\ref{subsec:uncertainties}), without which the profle asymmetry would be greatly overestimated.  Fitting the corrected distribution functions with a half-Gaussian, the standard deviation is $\sigma = 0.08$ for $A_F$ and $\sigma = 0.14$ for $A_C$. 
We define an object as asymmetric if its asymmetry parameters exceed $3\,\sigma$. By this criterion, the cumulative distribution functions yield an asymmetry fraction of $\sim 20\%$ (17.5\%$\pm$0.2\% for $A_F$ and 21.4\%$\pm$0.2\% for $A_C$; Figure~\ref{fig:asydistribution}b). The uncertainties for the fractions are 68.3\% confidence intervals calculated for a binomial distribution \citep{Cameron2011PASA...28..128C}. It is non-trivial to compare our results to those in the literature.  While our asymmetry fractions are roughly consistent with those quoted by \cite[][22\%]{Bournaud2005A&A...438..507B} and 
\cite[][9\%]{Espada2011A&A...532A.117E}, they are notably lower than the $\sim 50\%$ asymmetry fractions claimed by earlier studies that were based on less stringent classification criteria for profile asymmetry \citep{Richter1994A&A...290L...9R, Haynes1998AJ....115...62H, Matthews1998AJ....116.1169M}.  Most studies (but see \citealt{Watts2020MNRAS.492.3672W}) did not consider the systematic bias due to S/N.

We note that $A_C$ has a more prominent tail toward high values than $A_F$; the two parameters are significantly but not tightly correlated (Pearson's correlation coefficient $r = 0.37$).  They probe different and complimentary aspects of the \HI\ asymmetry.  Defined as the ratio of the integrated flux of the blueshifted and redshifted sides of the CoG, $A_F$ is sensitive to the asymmetry of the edges of the line profile.  On the other hand, $A_C$, defined as the ratio of the slopes of the rising portion of the CoG of the two sides of the line, is more sensitive to the asymmetry within the inner 85\% of the total gas distribution. The parameter $A_F$ gauges perturbations imparted by external, environmental effects, and $A_C$ should better reflect perturbations internal to the galaxy itself.  A forthcoming work (N. Yu et al., in preparation) will investigate the dependence of \HI\ asymmetry on the environmental and physical properties of galaxies.

\subsection{Statistical Properties of \HI\ Profile Shapes}

Figure~\ref{fig:shapedistribution} shows the distribution of profile shape parameters. The shape parameters $C_V$ and $K$ increase when a profile changes from double-horned to flat-topped and single-peaked, centered on a value of $C_V = 3.4$ or $K = 0$ for a flat-topped profile. Given the median uncertainties for $C_V$ (0.33) and $K$ (0.022), we classify the line profiles into three generic types by the following criteria: single-peaked ($C_V>3.7$, $K>0.02$), flat-topped ($3.1 \leq C_V\leq3.7$, $-0.02\leq K \leq 0.02$), and double-horned ($C_V<3.1$, $K<-0.02$). Example profiles are shown in Figure~\ref{fig:typical}.  According to the $C_V$ criteria, the fractions of single-peaked, flat-topped, and double-horned profiles in the ALFALFA main sample are $36.6 \pm 0.3\%$, $27.8 \pm 0.3\%$, and $35.6 \pm 0.3\%$, respectively.  The corresponding statistics adopting the $K$ criteria are $35.2 \pm 0.3\%$, $26.9 \pm 0.3\%$, and $37.9 \pm 0.3\%$, respectively. The two parameters give quite similar fractions for the three types of profile shape.  As $K$ makes use of the full information from the CoG, and not just the ratio of two line widths used to define $C_V$, the former is superior to the latter as a shape indicator, and therefore we consider the statistics based on $K$ to be more accurate. 

\begin{figure*}[!htbp]
\epsscale{1.1}
\plotone{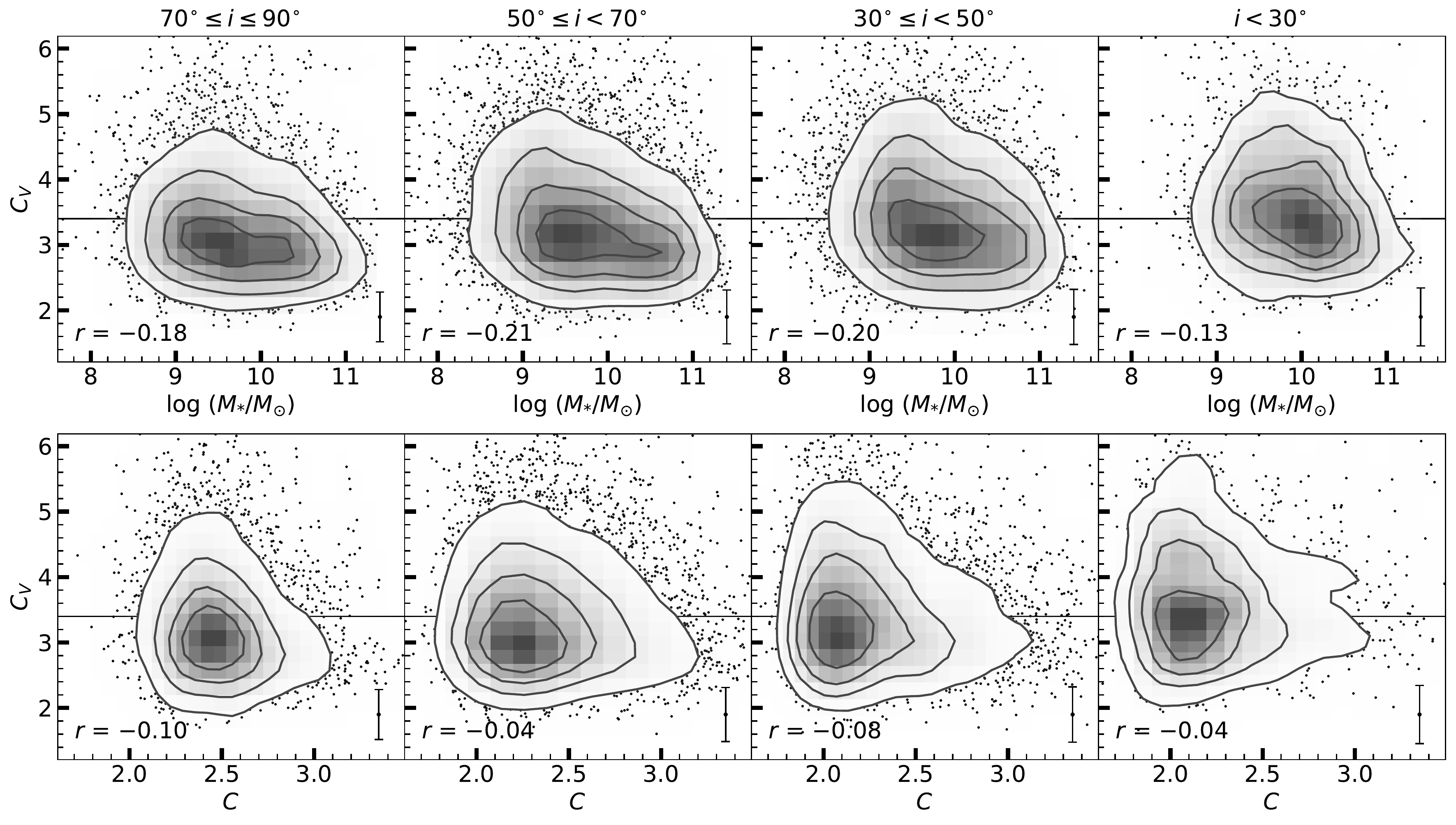}
\caption{The variation of profile concentration $C_V$ on (top) stellar mass $M_*$ and (bottom) optical concentration $C$ for subsamples binned by inclination angle: from left to right, $70^{\circ}\leq i\leq 90^{\circ}$, $50^{\circ}\leq i< 70^{\circ}$, $30^{\circ}\leq i< 50^{\circ}$, and $i< 30^{\circ}$. The density of points is highlighted by grayscale and contours.  The Pearson's correlation coefficient and a typical error bar for $C_V$ are given in the bottom of each panel. The contours in each panel show the levels of sample inclusion, which vary from 20\%, 40\%, 60\%, and 80\% from inside out.}
\label{fig:cvDep}
\end{figure*}

\begin{figure*}[!htbp]
\epsscale{1.1}
\plotone{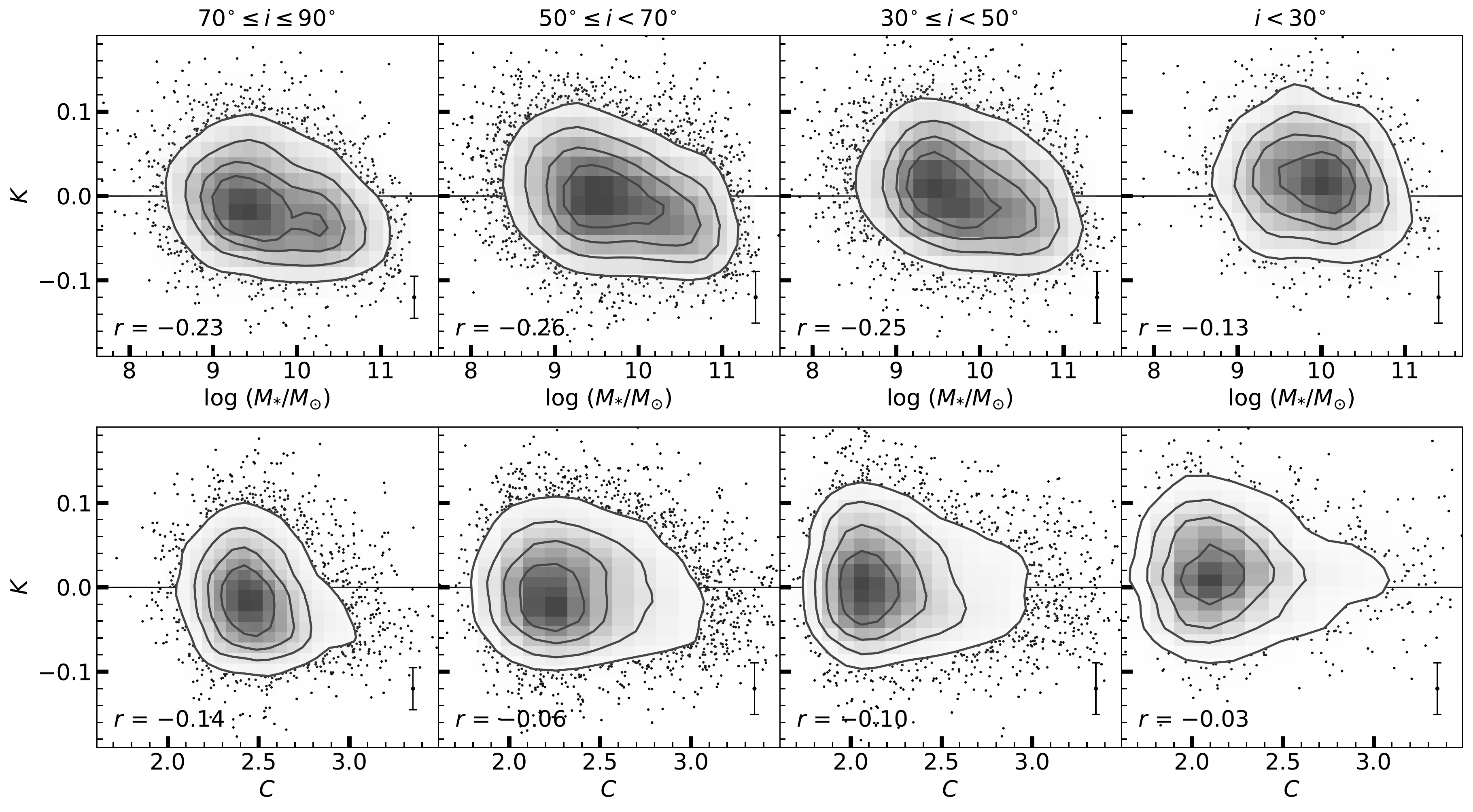}
\caption{The variation of profile shape $K$ on (top) stellar mass $M_*$ and (bottom) optical concentration $C$ for subsamples binned by inclination angle: from left to right, $70^{\circ}\leq i\leq 90^{\circ}$, $50^{\circ}\leq i< 70^{\circ}$, $30^{\circ}\leq i< 50^{\circ}$, and $i< 30^{\circ}$. The density of points is highlighted by grayscale and contours.  The Pearson's correlation coefficient and a typical error bar for $K$ are given in the bottom of each panel. The contours in each panel show the levels of sample inclusion, which vary from 20\%, 40\%, 60\%, and 80\% from inside out.}
\label{fig:kappaDep}
\end{figure*}

The observed line profile is affected by projection effects. An intrinsically double-horned profile, which is characteristic of most massive disk galaxies having a sufficiently extended \HI\ distribution that samples the flat part of the rotation curve (e.g., \citealt{Roberts1978AJ.....83.1026R, Martinsson2016AA...585A..99M}), would appear single-peaked when viewed face-on, with the width of the profile determined primarily by the turbulence motions of the \HI\ disk.  However, besides inclination angle, the \HI\ profile also imprints the velocity and spatial distribution of the gas \citep{El-Badry2018MNRAS.477.1536E, Yu2022}. Thus, while more face-on systems tend to have higher values of $C_V$ and $K$, such that the \HI\ profile is more likely to be single-peaked compared to edge-on systems (Figures~\ref{fig:shapedistribution}a and \ref{fig:shapedistribution}c), each of the bins of inclination angle covers a wide range of values for $C_V$ and $K$.

At a given inclination angle, galaxies with a higher stellar mass or optical concentration have a greater tendency to exhibit double-horned \HI\ profiles (lower $C_V$ or $K$) compared to galaxies of lower mass or concentration (Figures~\ref{fig:cvDep} and \ref{fig:kappaDep}). The formal statistical significance is not high, as judged by the Pearson's correlation coefficients, but the trends are persistent.  The overall dependence of profile shape on galaxy stellar mass resembles a similar trend seen in the study by \cite{El-Badry2018MNRAS.477.1536E}, which was based on a sample of $\sim 2000$ low-redshift, \HI-detected galaxies. These authors attributed the scatter in the relation between profile shape and galaxy stellar mass to 
inclination angle.  However, our results show that the scatter is still noticeable after fixing the inclination angle to a narrow range, suggesting that the scatter may be linked to physical factors, such as \HI\ content, \HI\ radial distribution, and velocity field. The trends with stellar mass are stronger than those with concentration, and they are more prominent for $K$ than for $C_V$. Moreover, the observed scatter of $C_V$ and $K$ at fixed $M_*$ or $C$ is substantial, and real, larger than the typical measurement uncertainties.  The intrinsic scatter can be used to diagnose physical processes that can lead the \HI\ profiles to depart from the expected norm.  For example, \cite{Zuo2022} note that gas-rich, major mergers, despite being massive galaxies, have a high frequency of single-peaked \HI\ profiles, likely a consequence of central gas inflow induced by gravitational torques.

This work represents the largest, most uniform, quantitative analysis of \HI\ profile shapes to date. Previous statistics on the \HI\ profile shape of nearby galaxies are not only highly limited, based on fragmentary samples of $\lesssim 100$ objects, but they also derive from subjective, visual classification (e.g., \citealt{Geha2006ApJ...653..240G, Ho2008ApJ...681..128H, Zhou2018PASP..130i4101Z}).

\section{Summary}
\label{sec:sum}

We apply the curve-of-growth technique of \citet{Yu2020ApJ...898..102Y} to perform a uniform analysis of the integrated \HI\ 21~cm spectra of 29,958 galaxies with unambiguous optical counterparts from the ALFALFA survey.  The main sample consists of nearby (median $z = 0.026$), gas-rich galaxies spanning a wide range of \HI\ mass (\MHI\ $\approx 10^{7.2}-10^{10.6}$ \msun).  Apart from basic parameters such as central velocity and total flux, the main catalog provides new measures of line width ($V_{85}$), profile asymmetry ($A_F$ and $A_C$), and line shape ($C_V$ and $K$).  We identify sources that may be confused by nearby, potentially physically associated neighbors, information that itself can be used to investigate the effect of local environment or dynamical interactions.  For the subsample of 13,511 galaxies that overlap with SDSS, which cover a wide range of stellar mass ($M_* \approx 10^{8.0}-10^{11.5}$ \msun, median $10^{9.7}$ \mstar), we provide higher level science products, including inclination angle-corrected line widths ($V_{85c}$), rotation velocities ($V_{\rm rot}$), and dynamical masses ($M_{\rm dyn}$), as well as ancillary physical properties (morphology, stellar mass, star formation activity), which will be used for a series of forthcoming applications. 

The availability of a homogeneous catalog of \HI\ parameters affords us the opportunity to reexamine the frequency of \HI\ profile asymmetry.  Across the full sample of galaxies under consideration, the fraction of galaxies with statistically significant \HI\ profile asymmetry is $\sim 20\%$.  The \HI\ spectra exhibit diverse profile shapes, but most can be classified as single-peaked ($35.2 \pm 0.3\%$), flat-topped ($26.9 \pm 0.3\%$), or double-horned ($37.9 \pm 0.3\%$). 
 The profile shapes reflect projection effects as well the intrinsic spatial and velocity distribution of the gas, which vary systematically with galaxy mass and mass distribution.  At a given inclination angle, double-horned profiles are preferentially associated with galaxies of higher stellar mass or optical concentration, while galaxies of lower mass or concentration tend to have single-peaked profiles. 

\acknowledgments
We thank the anonymous referee for very helpful comments and suggestions. This work was supported by the National Science Foundation of China (11721303, 11991052, 11903003, 12073002), the China Manned Space Project (CMS-CSST-2021-A04, CMS-CSST-2021-B02), and the National Key R\&D Program of China (2016YFA0400702). 
We thank Martha Haynes for kindly providing the spectra of ALFALFA survey. We are grateful to Pei Zuo, Yuming Fu, Jinyi Shangguan, Tianqi Huang, and Zhiwei Pan for useful advice and discussion. This research has made use of the NASA/IPAC Extragalactic Database ({\url http://ned.ipac.caltech.edu}), which is funded by the National Aeronautics and Space Administration and operated by the California Institute of Technology. We used Astropy, a community-developed core Python package for astronomy \citet{AstropyCollaboration2013A&A...558A..33A}.

\begin{rotatetable*}
\begin{deluxetable*}{crrD@{$\pm$}DD@{$\pm$}DD@{$\pm$}DD@{$\pm$}DD@{$\pm$}DD@{$\pm$}DD@{$\pm$}DD@{$\pm$}DrrD@{$\pm$}Dc}
\tablenum{2}
\centering
\addtolength{\tabcolsep}{-3.1pt}
\tabletypesize{\footnotesize}
\tablecolumns{15}
\tablewidth{0pt} 
\tablecaption{Parameters Derived from the \HI\ Spectra of the Main Sample}
\tablehead{
\colhead{Galaxy} &
\colhead{R.A.} &
\colhead{Decl.} &
\multicolumn4c{$D_L$} &
\multicolumn4c{$V_c$} &
\multicolumn4c{$F$} &  
\multicolumn4c{$V_{85}$} &
\multicolumn4c{$A_F$} &
\multicolumn4c{$A_C$} &
\multicolumn4c{$C_V$} &
\multicolumn4c{$K$} &
\colhead{S/N} &
\colhead{$\sigma_{\rm spec}$} &
\multicolumn4c{log \MHI} &
\colhead{Notes} \\ 
\colhead{} &      
\colhead{($^{\circ}$)} & 
\colhead{($^{\circ}$)} &  
\multicolumn4c{(Mpc)} &  
\multicolumn4c{(\kms)} & 
\multicolumn4c{(Jy\ \kms)} &
\multicolumn4c{(\kms)} &
\multicolumn4c{} &
\multicolumn4c{} & 
\multicolumn4c{} & 
\multicolumn4c{} & 
\colhead{} &  
\colhead{} &  
\multicolumn4c{(\msun)} &  
\colhead{}
\\
\colhead{(1)} &
\colhead{(2)} &
\colhead{(3)} &
\multicolumn4c{(4)} &
\multicolumn4c{(5)} &
\multicolumn4c{(6)} &
\multicolumn4c{(7)} &
\multicolumn4c{(8)} &
\multicolumn4c{(9)} &
\multicolumn4c{(10)} &
\multicolumn4c{(11)} &
\colhead{(12)} &
\colhead{(13)} &
\multicolumn4c{(14)} &
\colhead{(15)}  
}
\decimals % to align according to decimals (column identifier: D)
\startdata
AGC~000001 & 0.65667 & 16.65222 & 82.8 & 2.2 & 5832 & 3 & 2.96 & 0.46 & 154 & 11 & 1.01 & 0.08 & 1.14 & 0.13 & 5.02 & 0.36 & 0.106 & 0.017 & 21.2 & 2.5 & 9.68 & 0.07 & 1p \\
AGC~000004 & 0.73708 & 4.20889 & 118.0 & 2.3 & 8487 & 14 & 1.40 & 0.25 & 217 & 28 & 1.61 & 0.23 & 1.68 & 0.38 & 4.58 & 0.73 & 0.020 & 0.033 & 9.5 & 2.5 & 9.66 & 0.08 & 1 \\
AGC~000006 & 0.79042 & 21.95972 & 88.8 & 2.2 & 6576 & 9 & 1.17 & 0.21 & 227 & 28 & 1.32 & 0.19 & 1.08 & 0.27 & 5.00 & 0.81 & 0.012 & 0.034 & 7.2 & 2.8 & 9.34 & 0.08 & 2 \\
AGC~000007 & 0.79667 & 15.96500 & 155.2 & 2.2 & 11252 & 25 & 4.04 & 0.65 & 646 & 47 & 1.15 & 0.11 & 1.71 & 0.21 & 2.70 & 0.24 & $-$0.055 & 0.018 & 28.2 & 1.9 & 10.36 & 0.07 & 1 \\
AGC~000008 & 0.81167 & 16.14556 & 13.2 & 1.3 & 1048 & 3 & 22.42 & 3.55 & 426 & 26 & 1.02 & 0.07 & 1.06 & 0.11 & 2.92 & 0.20 & $-$0.045 & 0.015 & 157.3 & 2.0 & 8.96 & 0.11 & 1 \\
AGC~000010 & 0.83542 & 8.61861 & 165.4 & 2.1 & 11933 & 3 & 6.66 & 1.05 & 225 & 14 & 1.16 & 0.08 & 1.53 & 0.16 & 2.77 & 0.20 & $-$0.045 & 0.016 & 32.5 & 3.4 & 10.63 & 0.07 & 1 \\
AGC~000011 & 0.83958 & 22.10250 & 62.8 & 2.4 & 4443 & 4 & 2.33 & 0.37 & 211 & 13 & 1.22 & 0.09 & 1.06 & 0.12 & 2.73 & 0.22 & $-$0.010 & 0.018 & 19.2 & 2.1 & 9.34 & 0.08 & 1 \\
AGC~000012 & 0.83500 & 29.79722 & 95.0 & 2.4 & 6980 & 3 & 2.98 & 0.47 & 224 & 13 & 1.04 & 0.08 & 1.00 & 0.11 & 2.75 & 0.21 & $-$0.047 & 0.015 & 22.5 & 2.2 & 9.80 & 0.07 & 1 \\
AGC~000013 & 0.87167 & 27.35139 & 105.9 & 2.2 & 7764 & 7 & 0.81 & 0.15 & 170 & 24 & 1.02 & 0.16 & 1.15 & 0.29 & 5.19 & 0.83 & 0.140 & 0.035 & 5.8 & 2.9 & 9.33 & 0.08 & 2 \\
AGC~000014 & 0.89625 & 23.20028 & 98.6 & 2.2 & 7236 & 3 & 6.74 & 1.07 & 319 & 20 & 1.13 & 0.08 & 1.40 & 0.14 & 2.57 & 0.18 & $-$0.060 & 0.016 & 40.4 & 2.5 & 10.19 & 0.07 & 1 \\
AGC~000015 & 0.93625 & 4.29806 & 160.1 & 2.1 & 11558 & 7 & 3.27 & 0.52 & 494 & 30 & 1.15 & 0.10 & 1.16 & 0.14 & 2.90 & 0.23 & $-$0.034 & 0.017 & 18.4 & 2.3 & 10.30 & 0.07 & 1 \\
AGC~000016 & 0.95375 & 7.47861 & 74.1 & 2.4 & 5245 & 3 & 11.85 & 1.87 & 153 & 9 & 1.08 & 0.08 & 1.19 & 0.12 & 2.52 & 0.18 & $-$0.076 & 0.015 & 74.0 & 3.3 & 10.19 & 0.07 & 1 \\
AGC~000017 & 0.93000 & 15.21806 & 20.3 & 4.3 & 876 & 3 & 6.72 & 1.06 & 87 & 5 & 1.01 & 0.07 & 1.03 & 0.10 & 3.28 & 0.23 & $-$0.017 & 0.015 & 74.6 & 2.4 & 8.82 & 0.20 & 1 \\
AGC~000021 & 1.03708 & 7.37889 & 87.7 & 2.3 & 6202 & 3 & 5.01 & 0.79 & 163 & 10 & 1.19 & 0.09 & 1.26 & 0.14 & 2.89 & 0.21 & $-$0.044 & 0.015 & 27.9 & 3.4 & 9.96 & 0.07 & 1 \\
AGC~000022 & 1.02083 & 10.29417 & 105.6 & 2.2 & 7743 & 7 & 2.98 & 0.47 & 370 & 23 & 1.00 & 0.07 & 1.09 & 0.14 & 2.77 & 0.22 & $-$0.057 & 0.018 & 13.7 & 2.9 & 9.89 & 0.07 & 1 \\
AGC~000023 & 1.05417 & 10.79028 & 108.8 & 2.2 & 7969 & 7 & 2.33 & 0.36 & 370 & 25 & 1.20 & 0.10 & 1.83 & 0.23 & 2.84 & 0.35 & $-$0.049 & 0.018 & 16.9 & 2.0 & 9.81 & 0.07 & 1 \\
AGC~000024 & 1.06125 & 22.58778 & 48.1 & 9.1 & 4446 & 3 & 5.40 & 0.86 & 162 & 10 & 1.01 & 0.07 & 1.01 & 0.10 & 3.12 & 0.22 & $-$0.032 & 0.015 & 46.6 & 2.3 & 9.47 & 0.18 & 1 \\
AGC~000025 & 1.10458 & 6.17778 & 78.7 & 15.2 & 5081 & 3 & 3.79 & 0.60 & 257 & 16 & 1.13 & 0.08 & 1.26 & 0.14 & 2.52 & 0.19 & $-$0.060 & 0.016 & 27.4 & 2.3 & 9.74 & 0.18 & 1 \\
AGC~000026 & 1.10208 & 31.47194 & 42.1 & 8.3 & 4954 & 3 & 10.23 & 1.62 & 218 & 13 & 1.01 & 0.07 & 1.04 & 0.10 & 2.80 & 0.20 & $-$0.056 & 0.015 & 75.5 & 2.4 & 9.63 & 0.18 & 1 \\
AGC~000027 & 1.12125 & 5.84556 & 44.5 & 9.2 & 3112 & 3 & 13.51 & 2.14 & 193 & 12 & 1.01 & 0.07 & 1.05 & 0.11 & 2.73 & 0.19 & $-$0.058 & 0.015 & 138.8 & 1.9 & 9.80 & 0.19 & 1 \\
AGC~000030 & 1.13625 & 33.55917 & 68.7 & 4.2 & 4761 & 4 & 2.60 & 0.40 & 224 & 13 & 1.00 & 0.07 & 1.00 & 0.11 & 2.51 & 0.19 & $-$0.074 & 0.016 & 15.5 & 3.0 & 9.46 & 0.09 & 1 \\
\enddata
\tablecomments{Col. (1): Galaxy name from the Arecibo General Catalog (AGC). Cols. (2)--(3): Equatorial coordinates (J2000) of the optical counterpart assigned by ALFALFA. Col. (4): Luminosity distance from the ALFALFA catalog. Col. (5): Flux intensity-weighted central velocity. Col. (6): Total integrated flux of the \HI\ line. Col. (7): Velocity width measured at 85\% of the total line flux. Col. (8): Integrated flux asymmetry. Col. (9): Flux distribution asymmetry. Col. (10): Profile concentration. Col. (11): Profile shape. Col. (12): S/N of the profile. Col. (13): Noise level of the profile. Col. (14): \HI\ mass, following $M_{{\rm H~{\tiny I}}} = 2.36\times10^5 \ D_{L}^2\ F \, M_\odot$ \citep{Roberts1962}; for a distance uncertainty of 10\% and a flux uncertainty of 15\%, the typical uncertainty of log~\MHI\ is 0.11 dex. Col. (15): Notes: 
1 = code 1 detection; 
2 = code 2 detection; 
r = there are data points with spectral weight less than 0.5, which are located at the edge of the emission signal; 
c = the \HI\ emission overlaps with that of nearby companion(s);
p = the \HI\ emission is not confused even though there is at least a projected neighbor within the Arecibo beam.
(This table is available in its entirety in machine-readable form.)}
\label{tab:co-add}
\end{deluxetable*}
\end{rotatetable*}

\begin{rotatetable*}
\begin{deluxetable*}{crD@{$\pm$}DD@{$\pm$}DrrD@{$\pm$}DD@{$\pm$}DD@{$\pm$}DD@{$\pm$}DrrrD@{$\pm$}D}
\tablenum{3}
\centering
\small\addtolength{\tabcolsep}{-2pt}
\tabletypesize{\footnotesize}
\tablecolumns{14}
\tablewidth{0pt} 
\tablecaption{Properties of the SDSS Subsample}
\tablehead{
\colhead{Galaxy} &
\colhead{$z$} &
\multicolumn4c{$R_{90}$} &
\multicolumn4c{$R_{\rm H\, \textsc{i}}$} &
\colhead{$C$} &
\colhead{$b/a$} &
\multicolumn4c{$i$} &
\multicolumn4c{$V_{\rm rot}$} &
\multicolumn4c{log $M_*$} &
\multicolumn4c{log SFR} &
\colhead{log SFR$_{\rm in}$} &
\colhead{$D_{\rm n}4000$} &
\colhead{EW(H$\alpha$)} &
\multicolumn4c{log $M_{\rm dyn}$} 
\\ 
\colhead{} &      
\colhead{} & 
\multicolumn4c{(kpc)} &  
\multicolumn4c{(kpc)} &  
\colhead{} & 
\colhead{} &
\multicolumn4c{($^{\circ}$)} &
\multicolumn4c{(\kms)} &
\multicolumn4c{(\msun)} &
\multicolumn4c{(\msun yr$^{-1}$)} &
\colhead{(\msun yr$^{-1}$)} &
\colhead{} &
\colhead{(\AA )} &
\multicolumn4c{(\msun)} 
\\
\colhead{(1)} &
\colhead{(2)} &
\multicolumn4c{(3)} &
\multicolumn4c{(4)} &
\colhead{(5)} &
\colhead{(6)} &
\multicolumn4c{(7)} &
\multicolumn4c{(8)} &
\multicolumn4c{(9)} &
\multicolumn4c{(10)} &
\colhead{(11)} &
\colhead{(12)} &
\colhead{(13)} &
\multicolumn4c{(14)} 
}
\decimals % to align according to decimals (column identifier: D)
\startdata
AGC~000007 & 0.03742 & 16.0 & 1.1 & 49.5 & 4.1 & 2.8 & 0.4 & 72 & 5 & 332 & 45 & 11.342 & 0.014 & $-$0.039 & 0.117 & $-$2.002 & 1.97 & 1.41 & 12.1 & 0.2 \\
AGC~000043 & 0.01759 & 4.2 & 0.1 & 16.3 & 1.4 & 2.1 & 0.4 & 69 & 5 & 97 & 10 & 9.268 & 0.046 & $-$0.966 & 0.078 & $-$2.030 & 1.20 & 31.58 & 10.5 & 0.2 \\
AGC~000056 & 0.01809 & 7.0 & 0.1 & 21.2 & 4.5 & 2.0 & 0.7 & 42 & 5 & 138 & 16 & 9.814 & 0.029 & $-$0.605 & 0.048 & $-$2.239 & 1.34 & 5.98 & 11.0 & 0.2 \\
AGC~000180 & 0.03664 & 17.2 & 0.5 & 27.2 & 2.6 & 2.6 & 0.3 & 76 & 6 & 199 & 45 & 11.210 & 0.048 & 0.305 & 0.163 & $-$1.713 & 1.84 & 1.65 & 11.4 & 0.2 \\
AGC~000194 & 0.01827 & 17.1 & 0.4 & 15.6 & 1.5 & 3.2 & 0.7 & 49 & 5 & 356 & 71 & 10.787 & 0.004 & $-$1.249 & 0.033 & $-$2.326 & 1.95 & 3.88 & 11.7 & 0.2 \\
AGC~000219 & 0.01750 & 8.7 & 0.1 & 33.4 & 7.7 & 2.5 & 0.3 & 73 & 5 & 135 & 14 & 10.069 & 0.022 & $-$0.358 & 0.037 & $-$2.134 & 1.42 & 8.47 & 11.1 & 0.2 \\
AGC~000230 & 0.01771 & 8.8 & 0.1 & 43.3 & 3.7 & 2.1 & 0.8 & 35 & 5 & 116 & 15 & 10.281 & 0.033 & 0.214 & 0.053 & $-$1.428 & 1.24 & 20.03 & 11.1 & 0.2 \\
AGC~000233 & 0.01760 & 5.2 & 0.1 & 21.7 & 1.9 & 2.3 & 0.9 & 29 & 5 & 124 & 19 & 10.183 & 0.021 & $-$0.010 & 0.061 & $-$1.306 & 1.35 & 16.09 & 10.9 & 0.2 \\
AGC~000247 & 0.03725 & 12.9 & 0.6 & 43.3 & 3.5 & 2.5 & 0.2 & 79 & 6 & 151 & 17 & 10.327 & 0.016 & $-$0.036 & 0.031 & $-$1.114 & 1.32 & 9.91 & 11.4 & 0.2 \\
AGC~000252 & 0.01785 & 7.5 & 0.3 & 16.4 & 1.4 & 2.2 & 0.6 & 50 & 5 & 139 & 15 & 9.536 & 0.025 & $-$0.812 & 0.056 & $-$2.551 & 1.49 & 3.11 & 10.9 & 0.2 \\
AGC~000317 & 0.01788 & 6.6 & 0.1 & 12.0 & 1.0 & 2.1 & 0.8 & 41 & 5 & 24 & 3 & 9.147 & 0.042 & $-$0.836 & 0.053 & $-$2.332 & 1.28 & 10.12 & 9.2 & 0.2 \\
AGC~000401 & 0.01393 & 10.7 & 0.4 & 13.5 & 1.4 & 2.8 & 0.3 & 75 & 6 & 95 & 12 & 10.899 & 0.030 & 0.202 & 0.362 & $-$0.437 & 1.41 & 20.36 & 10.4 & 0.2 \\
AGC~000419 & 0.01461 & 12.4 & 0.2 & 21.0 & 2.3 & 1.9 & 0.7 & 49 & 5 & 139 & 15 & 10.531 & 0.017 & $-$0.164 & 0.012 & $-$0.608 & 1.22 & 36.25 & 11.0 & 0.2 \\
AGC~000450 & 0.01788 & 10.6 & 0.2 & 25.3 & 5.7 & 2.7 & 0.6 & 52 & 5 & 217 & 26 & 11.020 & 0.013 & $-$0.015 & 0.095 & $-$1.891 & 1.71 & 0.76 & 11.4 & 0.2 \\
AGC~000461 & 0.01392 & 6.8 & 0.1 & 30.2 & 6.8 & 2.4 & 0.6 & 52 & 5 & 170 & 18 & 10.521 & 0.024 & 0.207 & 0.065 & $-$1.282 & 1.41 & 8.13 & 11.3 & 0.2 \\
AGC~000463 & 0.01484 & 10.1 & 0.2 & 23.8 & 2.1 & 2.0 & 0.9 & 26 & 5 & 218 & 38 & 10.889 & 0.027 & 0.219 & 0.059 & $-$0.529 & 1.21 & 18.13 & 11.4 & 0.2 \\
AGC~000466 & 0.01814 & 6.5 & 0.1 & 17.4 & 4.0 & 2.4 & 0.4 & 65 & 5 & 121 & 13 & 9.730 & 0.028 & $-$0.647 & 0.051 & $-$2.074 & 1.44 & 2.58 & 10.8 & 0.2 \\
AGC~000507 & 0.01757 & 16.3 & 1.4 & 36.2 & 7.5 & 2.5 & 0.2 & 90 & 5 & 198 & 22 & 10.746 & 0.032 & $-$0.025 & 0.099 & $-$1.396 & 1.52 & 6.53 & 11.5 & 0.2 \\
AGC~000533 & 0.01825 & 7.4 & 0.1 & 21.2 & 4.4 & 2.0 & 0.3 & 72 & 5 & 143 & 15 & 9.849 & 0.024 & $-$0.619 & 0.056 & $-$2.188 & 1.40 & 5.45 & 11.0 & 0.2 \\
AGC~000611 & 0.01814 & 8.2 & 0.1 & 29.7 & 2.6 & 2.3 & 1.0 & 13 & 5 & 345 & 134 & 10.444 & 0.020 & 0.339 & 0.044 & $-$1.073 & 1.30 & 13.13 & 11.9 & 0.4 \\
\enddata
\tablecomments{Col. (1): Galaxy name. Col. (2): Optical redshift from SDSS. Col. (3): Radius containing 90\% of the Petrosian flux in the $r$ band.  Col. (4): \HI\, radius calculated from the \HI\, mass-size relation of \citet{Wang2016MNRAS.460.2143W}. Col. (5): Optical ($r$-band) concentration index $C$. Col. (6): Axis ratio in the $r$ band. Col. (7): Optical inclination angle. Col. (8): Rotation velocity. Cols. (9)--(10): Stellar mass and SFR from spectral energy distribution fitting \citep{Salim2018ApJ...859...11S}, scaled to the distance of ALFALFA. Cols. (11)--(13): SFR, $D_{\rm n}4000$, and EW(H$\alpha$) from the MPA-JHU catalog, calculated within the central 3$^{\prime \prime}$ fiber of SDSS. SFR$_{\rm in}$ is scaled to the distance of ALFALFA and converted to the initial mass function of Chabrier (2003). Col. (14): Dynamical mass within $R_{\rm H\ \textsc{i}}$.  (This table is available in its entirety in machine-readable form.)}
\label{tab:opt}
\end{deluxetable*}
\end{rotatetable*}

\vfill\eject

%\bibliographystyle{yahapj}
%\bibliographystyle{apj}
%\bibliography{HIref}{}

\end{CJK*}
\end{document}